# Time-domain observation of ballistic orbital-angular-momentum currents with giant relaxation length in Tungsten

Tom S. Seifert[1,2], Dongwook Go[3], Hiroki Hayashi[4,5], Reza Rouzegar[1,2], Frank Freimuth[3,6], Kazuya Ando[4,5,7], Yuriy Mokrousov[3,6], Tobias Kampfrath[1,2]

[1]Freie Universität Berlin, 14195 Berlin, Germany
[2]Fritz Haber Institut der Max-Planck-Gesellschaft, 14195 Berlin, Germany
[3]Forschungszentrum Jülich, 52425 Jülich, Germany
[4]Department of Applied Physics and Physico-Informatics, Keio University, Yokohama 223-8522, Japan
[5]Keio Institute of Pure and Applied Sciences, Keio University, Yokohama 223-8522, Japan
[6]Johannes-Gutenberg-Universität Mainz, 55099 Mainz, Germany
[7]Center for Spintronics Research Network, Keio University, Yokohama 223-8522, Japan

**Abstract**

The emerging field of orbitronics exploits the electron orbital momentum $L$. Compared to spin-polarized electrons, $L$ may allow magnetic-information transfera with significantly higher density over longer distances in more materials. However, direct experimental observation of $L$ currents, their extended propagation lengths and their conversion into charge currents has remained challenging. Here, we optically trigger ultrafast angular-momentum transport in Ni|W|SiO$_2$ thin-film stacks. The resulting terahertz charge-current bursts exhibit a marked delay and width that grow linearly with W thickness. We consistently ascribe these observations to a ballistic $L$ current from Ni through W with giant decay length ($\sim 80$ nm) and low velocity ($\sim 0.1$ nm/fs). At the W/SiO$_2$ interface, the $L$ flow is efficiently converted into a charge current by the inverse orbital Rashba-Edelstein effect, consistent with *ab-initio* calculations. Our findings establish orbitronic materials with long-distance ballistic $L$ transport as possible candidates for future ultrafast devices and an approach to discriminate Hall- and Rashba-Edelstein-like conversion processes.

**Introduction**

Spintronics research aims at utilizing the flow of spin angular momentum carried by electrons to transport information and eventually manipulate magnetic order [1]. Actually, electrons have two distinct channels of angular momentum: the electron spin $S$ and orbital angular momentum $L$. While $S$ is successfully exploited in the field of spintronics to transport information by $S$ currents and to convert the latter into detectable charge currents by spin-to-charge current conversion ($SC$C) [2], $L$ has only recently gained attention in the field of orbitronics. To make this fascinating concept compatible and competitive with conventional electronics [3, 4], the speed of spin-orbitronic functionalities needs to be scalable to terahertz (THz) rates [5].

A first key advantage of $L$ over $S$ is that $L$ can assume arbitrarily high values for one electron, which is interesting for efficient magnetic-order manipulation [1, 6, 7] by $L$-induced torques [8-12]. Second, $L$-to-charge current conversion ($LC$C) does not rely on spin-orbit interaction (SOI), which opens the orbitronic workbench to abundant light metals [13]. Finally, $L$-currents are predicted to propagate over increased lengths reaching almost 100 nm [14].

Recent studies provided strong indications of $L$ transport and charge-to-$L$-current conversion by the orbital Hall effect (OHE) in a thin layer of a paramagnetic material (PM). The $S$ or $L$ accumulation resulting from an in-plane charge current was interrogated by magnetooptic imaging [13] or by measuring the torque it exerted on the magnetization of an adjacent thin-film ferromagnetic material (FM) [1, 8-12, 14-24]. The FM was chosen to be either susceptible to $S$ (e.g., Ni$_{81}$Fe$_{19}$, CoFeB) or $L$ accumulation (e.g., Ni).

Unfortunately, it remains experimentally challenging to measure $L$ curents by $LC$C. First, it is difficult to distinguish $LC$C by the inverse OHE (IOHE) from $LC$C by an inverse orbital Rashba-Edelstein effect (IOREE) because both phenomena obey identical macroscopic symmetries. Second and for the same reason, IOHE and IOREE are difficult to separate from their $SC$C counterparts, i.e., from the inverse spin Hall effect (ISHE) and the inverse spin Rashba-Edelstein effect (ISREE) [25]. Previous work, however, indicates different spatial propagation and relaxation dynamics of $S$ and $L$ currents [8, 9, 14]. Therefore, an experimental approach such as THz emission spectroscopy [26, 27], which monitors currents with femtosecond resolution, is perfectly suited to access the possibly different ultrafast $L$-vs-$S$ propagation and conversion dynamics.

Here, we study ultrafast signatures of $S$ and $L$ transport from a FM into a PM that is launched by exciting FM|PM stacks with a femtosecond laser pulse. $LC$C and $SC$C in the PM is measured by monitoring the emitted THz pulse. Upon changing the FM from Ni to $Ni_{81}Fe_{19}$ (Py) and interfacing them with the PMs Pt, Ti and W, we find the same characteristic sign changes in the emitted THz field as in previous magnetotransport studies [9]. Consequently, we interpret our observations as signatures of ultrafast $LC$C and $SC$C. Remarkably, the emitted THz field from Ni|W is temporally strongly delayed and broadened relative to Ni|Pt. The bandwidth and amplitude of the underlying charge-current burst decreases with W thickness, whereas its delay increases linearly. We assign this observation to long-distance ballistic $L$ transport in W, which has a more than one order of magnitude larger relaxation length than $S$ transport. Specifically, our data and calculations suggest a dominant contribution to $LC$C through the IOREE at the $W/SiO_2$ interface. Interestingly, this effect is absent in Ni|Ti and attributed to a dominant bulk $LC$C by the IOHE. Our results may help establish an ultrafast experimental and theoretical methodology to extract the propagation dynamics of $L$ currents.

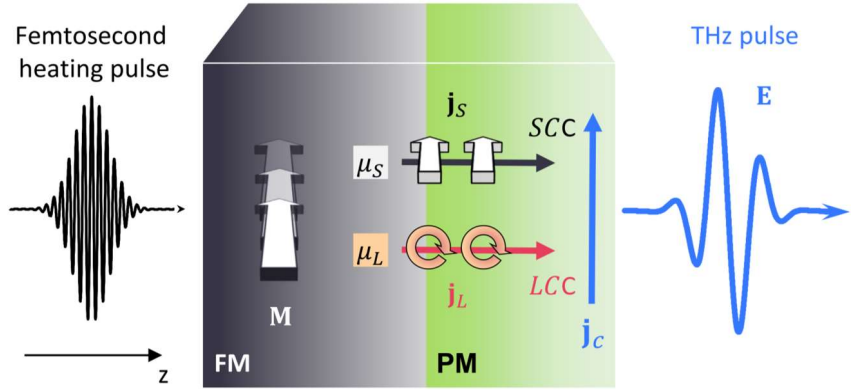

**FIGURE 1: Launching and detecting terahertz $S$ and $L$ currents.** Upon ultrafast laser excitation of the FM, an excess of FM magnetization **M** arises, leading to $S$ accumulation $\mu_S$, $L$ accumulation $\mu_L$ and the injection of spin and orbital currents $j_S$ and $j_L$, respectively, into the PM. Various bulk and interfacial $LC$C and $SC$C processes generate an ultrafast in-plane charge current $j_C$ that radiates a THz pulse with electric-field amplitude $E$ vs time $t$ directly behind the sample.

**Conceptual background.** Our approach is guided by the idea that $L$ currents have the same symmetry properties as $S$ currents, whereas $L$ transport is expected to have comparatively different spatiotemporal dynamics on ultrashort time and length scales [1, 8, 9, 14]. As schematically depicted in Fig. 1, a femtosecond optical pump pulse excites a FM|PM stack and triggers ultrafast $S$ and $L$ currents with density $j_S$ and $j_L$, respectively, from FM to PM. $SC$C and $LC$C result in ultrafast in-plane charge currents acting as sources of a THz electromagnetic pulse [28]. The resulting THz electric-field amplitude $E(t)$ directly behind the sample is proportional to the sheet charge current $I_C(t)$, i.e.,

$$E(t) \propto I_C(t) = \int_0^{d_{\text{FM}}+d_{\text{PM}}} dz \, [\theta_{LC}(z) j_L(z,t) + \theta_{SC}(z) j_S(z,t)]. \tag{1}$$

Here, $\theta_{LC}(z)$ and $\theta_{SC}(z)$ describe the local efficiency of instantaneous $LC$C and $SC$C, respectively. They include microscopic mechanisms like the ISHE or IOHE [27, 29], which occur in the bulk, or the ISREE and IOREE, which require regions of broken inversion symmetry such as interfaces [30, 31].

To understand the emergence of $j_S$ and $j_L$, we first note that sudden laser heating of the FM induces an $S$ accumulation $\mu_S$. The latter is proportional to the excess $S$ magnetization, i.e., the difference between the instantaneous $S$ magnetization and the equilibrium $S$ magnetization that would be attained at the instantaneous elevated electron temperature [32-35]. Consequently, the FM releases $S$ at a rate proportional to $\mu_S$, by transferring $S$ to both the crystal lattice through, e.g., spin flips, and the PM by, e.g., a flow $j_S \propto \mu_S$ of spin-polarized electrons. In complete analogy to $S$, we expect that laser heating also induces an $L$ accumulation $\mu_L$ that drives $L$ transport with density $j_L \propto \mu_L$. In other words, $\mu_S$ and $\mu_L$ are the driving forces of ultrafast $S$ and $L$ currents from FM to PM.

Recent studies on single-element FMs showed that the $S$- and $L$-type magnetizations exhibit very similar ultrafast time evolution following laser excitation [36-38]. Therefore, we expect a very similar time evolution of $\mu_S$ and $\mu_L$, i.e., $\mu_S(t) \propto \mu_L(t)$, where their amplitudes depend on details of the electronic structure [12]. Despite this common dynamic origin, the resulting $j_L(z,t)$ and $j_S(z,t)$ (Fig. 1) can have very different evolution because $S$ and $L$ may propagate differently through the FM/NM interface and the NM bulk.

Equation (1) does not account for contributions due to magnetic dipole radiation of the time-dependent magnetization and of photocurrents even in magnetic order, because both components can be discriminated experimentally [32, 39].

**Experiment details.** We study FM|PM thin-film stacks, where the two FMs Py and Ni are chosen for their high efficiency in generating $S$ and $L$ currents, respectively [9]. The PMs are chosen to have a strong ISHE (Pt, W) and IOHE (W, Ti) response. The reported signs for the ISHE are opposite for Pt vs W with a vanishing ISHE in Ti, but the expected IOHE signs are the same for all three PMs [40]. The studied FM|PM stacks have thicknesses of a few nanometers deposited onto 500 μm thick glass substrates or 625 μm thick thermally oxidized Si substrates (see Fig. S1 and Methods). The samples are characterized by optical and THz transmission spectroscopy [41], yielding the pump absorptance, DC conductivity and Drude relaxation rate (Fig. S2).

In our experiment (Fig. 1), ultrashort laser pulses (10 fs nominal duration, 800 nm center wavelength, 80 MHz repetition rate, 1.9 nJ pulse energy, 0.2 mJ/cm$^2$ incident fluence) derived from a Ti:sapphire oscillator excite the FM|PM samples. We record the emitted THz radiation by electrooptic sampling in a 1 mm or 10 μm thick ZnTe(110) or a 250 μm thick GaP(110) electro-optic crystal [42]. The resulting THz emission signal $S_{\text{THz}}(\mathbf{M}, t)$ vs time $t$ and sample magnetization $\mathbf{M}$ equals the THz electric-field waveform $E$ (Fig. 1) convoluted with a setup-response function [43]. The presented data is smoothed by convolution with a Gaussian (80 fs full width at half maximum) for better visibility unless noted otherwise.

All experiments are performed under ambient conditions unless stated otherwise. We apply an in-plane magnetic field of about 10 mT to the sample and monitor the THz field component perpendicular to $\mathbf{M}$. The component parallel to $\mathbf{M}$ is found to be minor (Fig. S3). Measurements with linearly and circularly polarized pump pulses reveal a negligible impact of the pump polarization on the THz emission (Fig. S4).

To isolate magnetic signals, we reverse $\mathbf{M}$ and focus on the odd-in-$\mathbf{M}$ THz signal $S_{\text{THz}}(t) = [S_{\text{THz}}(+\mathbf{M}, t) - S_{\text{THz}}(-\mathbf{M}, t)]/2$. Even-in-$\mathbf{M}$ signal components are more than two orders of magnitude smaller. As expected from a transport scenario, further experiments, in which the samples are reversed, reveal a dominant structural-inversion-asymmetry (SIA) character of the emitted THz signals. Minor contributions unrelated to SIA most likely arise from magnetic-dipole radiation due to ultrafast demagnetization (Fig. S5) [32].

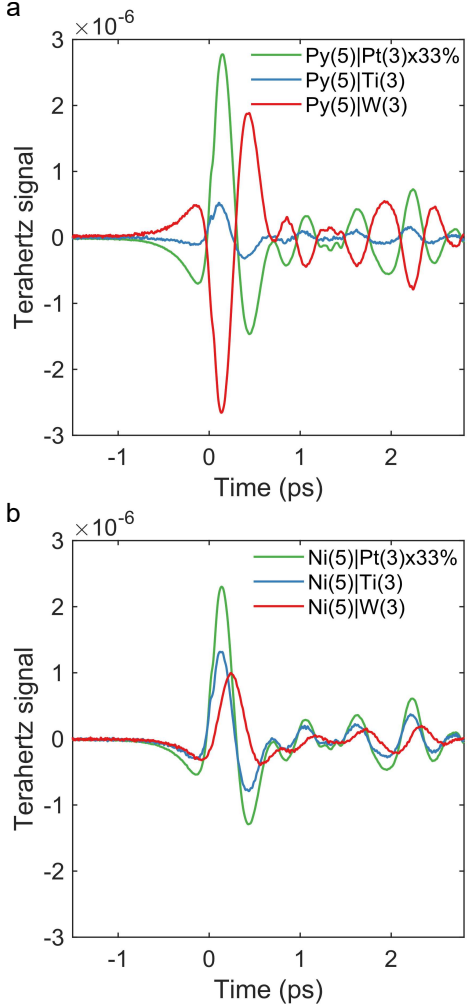

**FIGURE 2: Terahertz raw data.** THz emission signals $S_{\mathrm{THz}}(t)$ from FM|PM stacks with **a**, FM=Py and **b**, FM=Ni capped with PM=Pt, W or Ti. Note the rescaling of the Pt-based sample signals. Film thicknesses in nanometers are given as numerals in parenthesis. As THz detector, a 1 mm thick ZnTe(110) crystal was used.

**Results**

**FM=Py.** Figure 2a shows THz emission signals $S_{\mathrm{THz}}$ from Py|PM samples with PM=Pt, W, Ti, where the time-axis origin is the same for all signals. All three waveforms have identical shape. Minor differences in the shape of $S_{\mathrm{THz}}^{\mathrm{Py|Ti}}$ vs $S_{\mathrm{THz}}^{\mathrm{Py|Pt}}$ are attributed to contributions unrelated to SIA (see above and Fig. S6).

The relative signal magnitudes as well as the opposite polarities for PM=Pt and W are consistent with previous reports of ISHE-dominated THz emission [28]. The polarity of the signal from Py|Ti is the same as from Py|Pt and consistent with calculations and measurements, which found the same sign of the ISHE in Pt and the IOHE in Ti [13, 27, 40]. However, the Py|Ti signal has a significantly smaller amplitude than the Py|Pt signal even though Ti has a sizeable $LC$C efficiency. We ascribe this observation to a small amplitude of the $L$ current injected into Ti, consistent with the small $L$ component of the Py magnetization [9].

To summarize, for Py|PM, our THz signals are consistent with the notion that we observe transport of predominantly $S$ and $L$ into the PM bulk and its conversion into a charge current through the ISHE and the IOHE, respectively. At the FM/PM interface, a possible Rashba- or skew-scattering-type $LC$C or $SC$C [44] may make an additional yet relatively small contribution.

**FM=Ni.** When the FM=Py is replaced by Ni, the signal polarity remains the same for Pt and Ti, and the two waveforms for the different FMs exhibit identical dynamics (Fig. 2b and Fig. S7). In stark contrast, however, the signal polarity for Ni|W reverses with respect to Py|W, the waveform is less symmetric, and its maximum appears later than for Py|W. This striking observation indicates that Py|W and Ni|W show competing photocurrent-generation mechanisms, the dominance of which depends sensitively on the FM material. To gain more insight into the different dynamics in Ni|W, we next vary the W thickness.

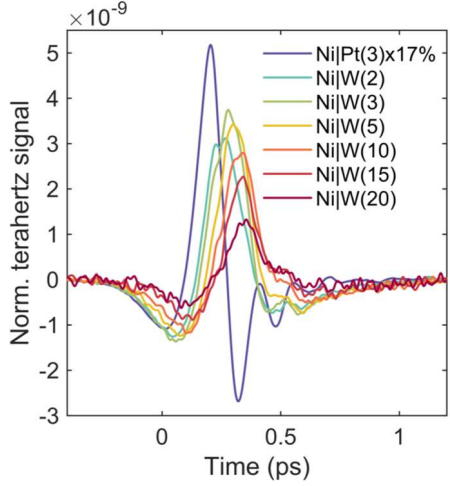

**FIGURE 3: Impact of W thickness in Ni|W.** THz emission signals for Ni|W samples with varying W thickness normalized to the absorbed pump-pulse fraction in the Ni layer and to the sample impedance (see Methods and Table S1). Note the rescaling of the reference signal from Ni|Pt. Film thicknesses in nanometers are given as numerals in parenthesis, except for Ni, which was always 5 nm thick. A 250 μm thick GaP(110) crystal was used as THz detector.

**Impact of W thickness.** Figure 3 shows THz emission signals from Ni|W($d_W$) for various W thicknesses $d_W$ and from a Ni|Pt reference sample. Consistent with Fig. 2b, we see a clear trend with increasing $d_W$ relative to Ni|Pt: The THz signal amplitude has the same sign, reduces with increasing $d_W$ and undergoes a significant reshaping from asymmetric (Ni|Pt) to more symmetric (Ni|W) around the signal maximum. Interestingly, $d_W = 2$ nm is already sufficient to induce a shift of the THz-signal maximum by about 100 fs with respect to Ni|Pt.

We emphasize that the changes in THz-signal dynamics solely originate from changing the PM thickness. Therefore, the FM is not primarily responsible for the signal-dynamics changes and, thus, considered as an PM-independent $S$ and $L$ injector in the following.

**Current dynamics in Ni|W.** To obtain a sample-intrinsic measurement of the $L$ transport and conversion dynamics, we extract the sheet charge current $I_C(t)$ flowing in Ni|W (Eq. (1)) normalized to the absorbed laser fluence in the Ni layer. This procedure eliminates any impact of sample exchange on pump-pulse absorption efficiency, sample impedance or setup response function (see Methods).

Figure 4a presents $I_C(t)$ in Ni|W with a resolution of 50 fs for various W thicknesses $d_W$. The $I_C(t)$ traces have striking features. (i) They exhibit the same polarity as Ni|Pt. (ii) The $I_C$ amplitude decreases approximately linearly with $d_W$ to about 50% after 20 nm (Fig. 4c), indicating attenuation and dispersion upon propagation. (iii) Their maximum shifts by delays $\Delta t_{\max} \propto d_W$ at a rate $\Delta t_{\max}/d_W \approx 4$ fs/nm (Fig. 4b), implying a velocity of 0.25 nm/fs. (iv) The $I_C$ width increases linearly at a rate of $\approx 8$ fs/nm (Fig. 4d). (v) The time-integrated current $\int dt\, I_C(t)$ only weakly decreases with $d_W$, thereby indicating an extremely large relaxation length >20 nm (Fig. 4e) [8, 9, 14].

Feature (i) implies that $I_C(t)$ cannot arise from $S$ transport. The reason is that $SC$C in Pt and W is dominated by the ISHE, the strength of which has an opposite sign in Pt and W according to theory [28, 45] and experiment [28]. This conclusion is strongly supported by feature (ii) because an $S$ current in W would relax over distances much smaller than 20 nm [41, 46]. Our data, therefore, strongly indicate that $L$ transport plus $LC$C is the dominant origin of the THz charge current in Ni|W.

Features (iii) and (iv) are a hallmark of a signal arising from ballistic-like transport of a pulse that is detected in an arrival layer. In this picture, the increase of the $I_C(t)$ width with $d_W$ arises from velocity dispersion along the $z$-direction of the particles that make up the pulse (Fig. 5a). Feature (v) implies a minor $LC$C in the W bulk because it would otherwise result in an integrated charge current $\int dt\, I_C(t)$ that increases monotonically with $d_W$.

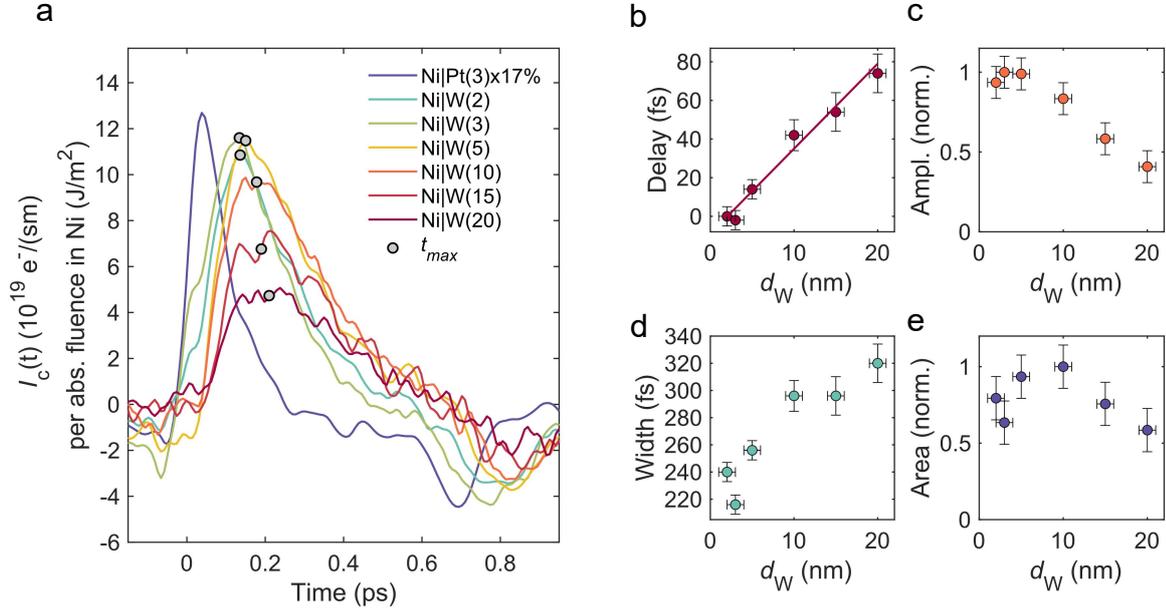

**FIGURE 4: Ultrafast charge currents in Ni|W. a**, Charge sheet currents $I_C(t)$ in Ni|W for various W thicknesses $d_W$ as extracted from the data of Fig. 3. The feature at 0.8 ps is a remainder of a THz-field reflection echo in the 10 μm ZnTe electro-optic detection crystal (see Methods). Film thicknesses in nanometers are given as numerals in parenthesis, except for Ni, which was always 5 nm thick. Note the rescaling of the Pt-based sample signal. The apparent signal delays and amplitudes are highlighted by a circular marker. **b**, Extracted time delay relative to the Ni(5)|W(2) sample with a straight line as a guide to the eye, **c**, relative amplitude at the delay marked in panel a, **d**, temporal width at half maximum, and **e**, integrated charge current between 0 to 0.7 ps vs $d_W$ from the data in panel a.

**Model: $L$ current and IOREE in Ni|W.** The preceding discussion suggests the following transport scenario in Ni|W. Upon excitation of the Ni layer, a transient $S$ and $L$ accumulation is induced (Fig. 1). Their dynamics are expected to be very similar (see above) [36-38] and monitored well by the ISHE charge current in Ni|Pt (Fig. 4a). The $L$ accumulation launches $L$ transport into W, which is carried by electron wavepackets, i.e., coherent superpositions of Bloch states such that $L$ is nonzero [1] (Fig. 5a). Finally, $LC$C is dominated in regions close to the W/SiO$_2$ interface. This interpretation is strongly supported by our *ab initio* calculations (see Methods and Figs. S8-S10). The calculated $L$ velocity (~ 0.1 nm/fs, Fig. S9) agrees well with our measurements (Fig. 4b). The calculated $LC$C (Fig. S10) reveals a giant interfacial $LC$C response in a thin W film with the same sign as the ISHE in Pt (Fig. 4a). An efficient interfacial $LC$C was already invoked in previous works [7, 24, 47-51].

The suggested scenario can explain all charge-current features (i)-(v) (Fig. 4) and is consistent with the above experimental findings. As the $j_L$ pulse propagates predominantly ballistically, its arrival in the W/SiO$_2$ $LC$C region is delayed by a time $\Delta t_{max} \propto d_W$. During propagation through PM=W, the $j_L$ pulse disperses due to different electron velocity projections onto the $z$-axis (Fig. 5a) and attenuation with a relaxation length >20 nm.

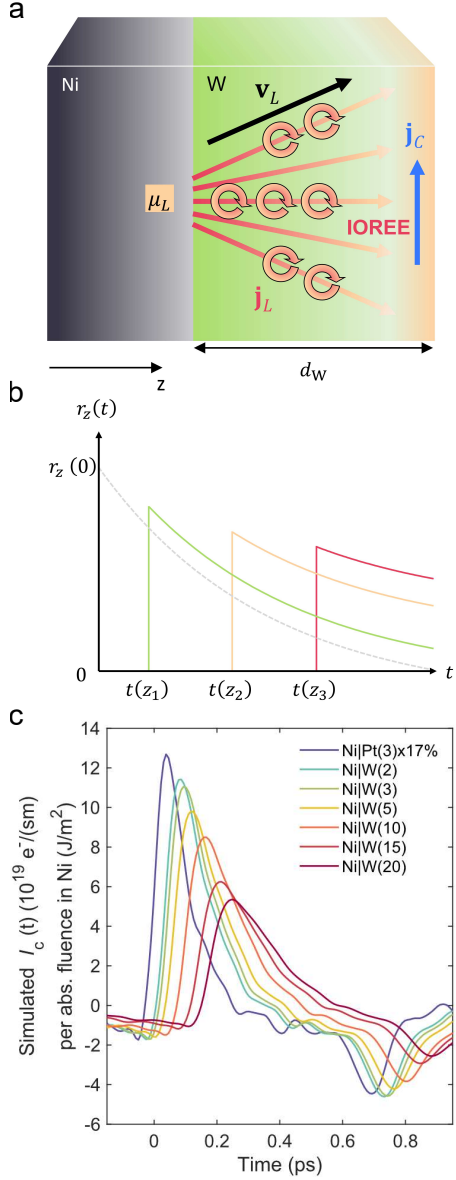

**FIGURE 5: Model of $L$ transport and IOREE in W. a**, Schematic of the suggested scenario for $L$ transport and $LC$C by the IOREE in Ni|W. The impulsive $L$ accumulation $\mu_L$ launches $L$-wavepackets of various velocities and, thus, an orbital current $\mathbf{j}_L$ into the W layer. Upon reaching the W back surface, $\mathbf{j}_L$ is converted into a transverse charge current $\mathbf{j}_C$ by the inverse orbital Rashba-Edelstein effect (IOREE). In the experiment, many of the point-like sources of orbital currents are superimposed along the Ni/W interface. **b**, Qualitative ballistic current densities $r_z(t)$ in response to a fictitious $\delta(t)$-like $\mu_L$ at different positions $z_1 < z_2 < z_3$ in W. **c**, Simulated IOREE charge currents $I_C(t)$ obtained by Eq. (2) in the case of ballistic transport. Parameters are a $L$-wavepacket velocity (panel a) $v_L = 0.14$ fs/nm, an $L$ decay length of 80 nm and a global scaling factor.

To model the charge-current dynamics in Ni|W (Fig. 3), we assume $j_L$ is solely driven by $\mu_L$ and, thus, given by the linear-response relationship

$$j_L(z,t) = (r_z * \mu_L)(t) = \int d\tau\, r_z(t-\tau)\mu_L(\tau). \tag{2}$$

Here, $r_z$ is the $L$-current density at position $z$ following a fictitious $\delta(t)$-like $L$ accumulation in Ni, and $\mu_L(\tau) \propto \mu_S(\tau)$ is given by the charge current measured in Ni|Pt (Fig. 4a). Assuming that the IOREE response at $z = d_W$ is instantaneous within the time resolution of our experiment, the total charge current $I_C(t)$ is proportional to $j_L(d_W, t)$.

In the Methods section, we analytically calculate the response function $r_z$ for the cases of purely ballistic (Eqs. (9), (10) and Fig. 5b) and purely diffusive transport (Eq. (12)). For the ballistic case, the modeled $I_C(t)$ curves (Fig. 5c) reproduce the measured charge currents in Ni|W (Fig. 4a) semiquantitatively for a relaxation length of 80 nm and a dominant $L$ wavepacket velocity of $v_L = 0.14$ nm/fs. These values agree well with the estimates obtained above (Figs. 4b-e) and with the orbital velocity obtained from our *ab-initio* calculations (see Fig. S9).

We note that even better agreement of modeled and calculated $I_C(t)$ could be obtained by considering different distributions of the $L$-velocity directions (Fig. 5a and Eq. (9)). In contrast, for diffusive transport, our model reproduces the experimental data less favorably (Fig. S11). We conclude that the observed currents (Fig. 4) have a significant ballistic component.

To summarize, the THz charge currents in Ni|W (Fig. 4) can be considered as signatures of $L$ currents injected into W. The charge-current generation [see Eq. (1)] is dominated by an extremely long-range $j_L$ and $LC$C at the W/SiO$_2$ interface, i.e., by $\theta_{LC}$ at $z = d_{\text{FM}} + d_{\text{PM}}$. Such long-range angular momentum transport is a unique feature of orbitronic materials, first indications of which were found previously in Ti [8, 9].

**Discussion**

Our interpretation neglects other possible contributions to the photocurrent $I_C(t)$. First, the inverse Faraday effect as a source of $S$ and $L$ currents can be ruled out by the pump-polarization independence (see Fig. S4).

Second, for the $S$ channel, a dominant Seebeck-type contribution due to an electronic temperature difference $\Delta T_{\text{FM/PM}}$ across the Ni/PM interface is neglected as found in previous studies [32]. For the $L$ channel, we estimate $\Delta T_{\text{Ni/PM}}$ right after pump pulse absorption (see Methods) and find $\Delta T_{\text{Ni/Pt}} \sim +400$ K and $\Delta T_{\text{Ni/PM}} \sim -100$ K in Ni|Ti and Ni|W. The observed THz-emission signals, in contrast, show the same sign for all three samples (Fig. 2b). Therefore, interfacial electronic temperature differences are a minor driving force and have minor impact on the $L$ and $S$ transport. Pump-propagation simulations also show that, even for the thickest samples, pump-intensity gradients in the FM and PM bulk are relatively small (Fig. S12).

Third, regarding transport in W, we consider dominant angular-momentum transport by magnons unlikely because W is not magnetically ordered. Acoustic phonons are excluded because sound velocity in W is < 0.01 nm/fs [52] and, thus, significantly slower than the observed transport velocity. An outstandingly long propagation of $S$ transport is ruled out, too, because the Drude scattering times for all studied samples are substantially shorter ( ≪ 50 fs, Fig. S2) than the peak delays of $I_C(t)$ (Fig. 4b). This argumentation also implies different dissipation mechanisms for $L$ and $S$ currents, which, however, require further investigations.

Fourth, even though our data imply a dominant IOREE contribution to charge-current generation, the positive shoulder-like feature at around time zero for $d_W \leq 3$ nm in Fig. 4a may indicate a small contribution of bulk $LC$C, i.e., the IOHE, which would agree with the sign of the IOHE in W. A $L$-to-$S$ conversion plus ISHE in the PM [23] might contribute but is considered negligible here given the good agreement of our experimental data (Fig. 4) and the IOREE scenario (Fig. 5). The dominance of an $L$-type angular momentum current in Ni|W highlights the role of Ni as an $L$ source and indicates that the Ni/W interface may transmit $L$ currents more efficiently than $S$ currents.

We finally turn to other interesting aspects of our study. A more detailed comparison of Fig. 2a and 2b reveals further changes in amplitude between Ni- and Py-based samples. The pronounced amplitude changes for PM=W or Pt when changing FM=Py to Ni are related to the intricate interplay of all parameters in Eq. (1) in addition to changes in the relative amplitudes of $\mu_S$ and $\mu_L$, and interface transmission coefficients for $j_S$ and $j_L$. Therefore, further experiments for a robust separation of $S$ and $L$ transport are needed.

We further find that the THz signals from the Ni-based samples increase linearly with pump fluence. Slight sublinearities at the highest fluences do not alter the THz emission dynamics (Fig. S13). We emphasize that samples deposited on Si rather than glass substrates show very similar THz emission characteristics (Fig. S1). These observations demonstrate the robustness of the observed effects.

When adding a Cu layer on top of the Ni|W sample, we find similar THz emission signals (Fig. S14). Future studies are needed to elaborate the exact character of the IOREE for different interfaces. Interestingly, a Cu intermediate layer in Ni|Cu|W slightly modifies the amplitude and dynamics of the THz signal, suggesting that Cu does not block $L$ transport strongly.

Regarding earlier reports of different THz emission dynamics in Fe|Au and Fe|Ru samples [29], we note that a possible IOREE in Ru can, in hindsight, not be excluded. A dominant IOREE might also explain the seemingly strong dependence of the Fe|Ru THz emission dynamics on the exact growth details [29, 53, 54].

In conclusion, we observe THz-emission signals from optically excited Ni|W stacks that are consistently assigned to ultrafast injection of $L$ currents into W and long-distance ballistic transport through W. Remarkably, we find strong indications for the occurrence of the IOREE in both experiment and theory. This result can be considered as time-domain evidence of the long-range nature of orbital currents and the IOREE in typical metals such as W.

Our study highlights the power of broadband THz emission spectroscopy in disentangling spin and orbital transport and Hall- and Rashba-Edelstein-like angular-momentum conversion processes based on their dynamic signatures. We find that Py vs Ni are, respectively, attractive $S$ and $L$ sources, whereas Pt vs W are, respectively, good $S$- and $L$-to-charge converters with distinctly different efficiency and dynamics for $S$ vs $L$. We believe that our results are a significant step toward the identification of ideal sources and detectors of either $S$ or $L$ currents. State-of-the-art theoretical predictions of $SC$C and $LC$C efficiencies will play an important role in this endeavor.

## Methods

**Current extraction.** To extract the in-plane sheet current flowing inside the sample from the measured THz signal $S_{\text{THz}}$, we first measure our setup response function $H_{SE}$ by having a reference electro-optic emitter (50 µm GaP on a 500 µm glass substrate) at the same position as the sample, which yields a reference THz signal $S_{\text{THz}}^{\text{ref}}$. By calculating the emitted THz electric field from that reference emitter $E^{\text{ref}}$, $H_{SE}$ is determined by solving the convolution $S_{\text{THz}}^{\text{ref}}(t) = (H_{SE} * E^{\text{ref}})(t)$ for $H_{SE}$ [43]. Further measured inputs for this calculation are the excitation spot size with a full width at half maximum of 22 µm, the excitation pulse energy of 1.9 nJ and a transform limited pump pulse with a spectrum centered at 800 nm and 110 nm full width at half maximum. We perform the deconvolution directly in the time domain by recasting it as a matrix equation [55].

Next, the electric field $E$ directly behind the sample is obtained from the recorded THz signal $S_{\text{THz}}$ with the help of the derived function $h$ by solving again the similar equation $S_{\text{THz}} = (H_{SE} * E)(t)$ for $E$. Finally, the sheet charge current (see Table S1) as shown in Fig. 3 is derived from a generalized Ohm's law [28], which in the frequency domain at frequency $\omega/2\pi$ reads

$$E(\omega) = eZ(\omega)I_C(\omega). \qquad (3)$$

Here, $-e$ is the electron charge and the sample impedance $Z(\omega)$ is given by $Z_0/[1 + n_{\text{sub}} + Z_0 d\sigma(\omega)]$ with the free-space impedance $Z_0$, the substrate refractive index $n_{\text{sub}} \approx 2$ [56] and the metal-stack thickness $d$. The measured mean sample conductivity $\sigma$ (see Table S1) is assumed to be frequency-independent due to the large Drude scattering rate (see Fig. S2). To enable comparison of THz currents from different samples, we normalize $I_C$ by the absorbed fluence in the FM layer. The data shown in Fig. 3 was obtained in a dry-air atmosphere.

**Sample preparation.** The FM|PM samples (FM = Ni and Py, PM = Pt, Ti, Cu and W) are fabricated on glass substrates of 500 µm thickness or thermally oxidized Si substrates of 625 µm thickness by radio frequency magnetron sputtering under 6N-purity-Ar atmosphere. The sample structure and thickness are described in Table S1. For the sputtering, the base pressure in the chamber is lower than 5.0×10⁻⁷ Pa. To avoid oxidation, 4-nm-thick $SiO_2$ is sputtered on the surface of the films. All sputtering processes are performed at room temperature. The W films are predominantly in the $\beta$-phase for $d_W < 10$ nm with an $\alpha$-phase content that grows with $d_W$ and dominates for $d_W > 10$ nm [9].

**Estimate of electronic temperatures.** We calculate the electronic temperature increase $\Delta T_{e0}$ upon pump-pulse absorption by

$$\Delta T_{e0} = \sqrt{T_0^2 + \frac{2F_l}{d\gamma}} - T_0. \qquad (4)$$

Here, $T_0 = 300$ K is the ambient temperature, $F_l$ is the absorbed fluence in the respective layer $l = $ FM or PM (see Table S1), $d$ is the layer thickness, and $\gamma T_e$ is the specific electronic heat capacity with $\gamma = 300$ J/m³ K² for W, 320 J/m³ K² for Ni and 330 J/m³ K² for Ti and 90 J/m³ K² for Pt [57]. To obtain the absorbed fluences in each layer, we note that the pump electric field is almost constant throughout the sample (see Fig S8). Therefore, local pump absorption scales solely with the imaginary part Im $\varepsilon$ of the dielectric function at a wavelength of 800 nm, which equals 22.07 for Ni, 9.31 for Pt, 19.41 for Ti and 19.71 for W [58]. Consequently, the absorbed fluence is determined by

$$F_l = F_{\text{tot}} \frac{d_l \, \text{Im} \, \varepsilon_l}{d_{\text{FM}} \, \text{Im} \, \varepsilon_{\text{FM}} + d_{\text{PM}} \, \text{Im} \, \varepsilon_{\text{PM}}} \qquad (5)$$

with the total absorbed fluence $F_{\text{tot}}$ that is obtained from the absorbed pump power (see Table S1) and the beam size on the sample (see above).

**Experimental error estimation.** The error bars for the delay of the transient charge current $I_C(t)$ (Fig. 4b) are estimated as ±20% of the read-off delay value (circular markers in Fig. 4a), but no less than 5 fs. The uncertainty in the relative amplitude (Fig. 4c) and relative area (Fig. 4e) is estimated as, respectively, ±20% and ±10%. The latter two uncertainty estimates reflect the typical signal-to-noise ratio of the extracted current traces (Fig. 4a). The error bars for the width of $I_C(t)$ (Fig. 4d) are obtained from the uncertainty of the delay (Fig. 4b), with subsequent multiplication by $\sqrt{2}$, which accounts for error propagation of a difference of two quantities.

**Model of $L$ transport.** To model the ballistic current in the PM, we assume that a $\delta(t)$-like transient $L$ accumulation in the FM generates an electronic wavepacket, which has orbital angular momentum $\Delta L_{\mathbf{k}0}$ along the direction of the F magnetization $\mathbf{M}$ and mean wavevector $\mathbf{k}$ in the PM right behind the FM/PM interface at $z = 0^+$ (Fig. 5a).

In the case of purely ballistic transport, this wavepacket propagates into the PM bulk according to $\Delta L_{\mathbf{k}}(z,t) = \Delta L_{\mathbf{k}0} \delta(z - v_{\mathbf{k}z} t)$, where $v_{\mathbf{k}z}$ is the $z$ component of the wavepacket group velocity. Note that we restrict ourselves to $\mathbf{k}$ with nonnegative $v_{\mathbf{k}z}$. The total pump-induced $L$-current density flowing into the depth of the PM layer is for $z > 0$ given by the sum

$$r_z(t) = \sum_{\mathbf{k},\, v_{\mathbf{k}z} \geq 0} \Delta L_{\mathbf{k}0} v_{\mathbf{k}z} \delta(z - v_{\mathbf{k}z} t). \tag{6}$$

Assuming that $\Delta L_{\mathbf{k}0}$ arises from states not too far from the Fermi energy, the summation of Eq. (6) is approximately proportional to an integration over the Fermi-surface parts with $v_{\mathbf{k}z} \geq 0$. One obtains

$$r_z(t) = e^{-t/\tau} \int_0^\infty dv_z\, w(v_z) v_z\, \delta(z - v_z t), \tag{7}$$

where $z > 0$, and

$$w(v_z) = \sum_{\mathbf{k},\, v_{\mathbf{k}z} \geq 0} \Delta L_{\mathbf{k}0} \delta(v_{\mathbf{k}z} - v_z) \tag{8}$$

is the $L$-weight of the $z$-axis group velocity $v_z$. In Eq. (7), we phenomenologically account for relaxation of the ballistic current with time constant $\tau$ by introducing the factor $e^{-t/\tau}$. Performing the integration of Eq. (7) yields

$$r_z(t) = \frac{e^{-t/\tau}}{t} \frac{z}{t} w\left(\frac{z}{t}\right). \tag{9}$$

To determine a plausible shape of $w(v)$, we note that $\Delta L_{\mathbf{k}0}$ is nonzero within several 0.1 eV around the Fermi energy $E_F$ owing to the width of the photoexcited and rapidly relaxing electron distribution [32]. Assuming a spherical Fermi surface and isotropic $\Delta L_{\mathbf{k}0}$, we have $\Delta L_{\mathbf{k}0} \propto \delta(E_\mathbf{k} - E_F)$ and $v_{\mathbf{k}z} = v_F \cos\theta$, where $E_\mathbf{k}$ is the band structure, $v_F$ is the Fermi velocity, and $\theta$ is the angle between $\mathbf{k}$ and $z$ axis. Therefore, after turning Eq. (8) into an integral, the integrand $\Delta L_{\mathbf{k}0} \delta(v_F \cos\theta - v_z)\, d^3\mathbf{k}$ becomes proportional to $\delta(v_F \cos\theta - v_z)\, d\cos\theta$ in spherical coordinates, leading to

$$w(v_z) \propto \int_0^1 d\cos\theta\; \delta(v_F \cos\theta - v_z) \propto \Theta(v_F - v_z). \tag{10}$$

In other words, all velocities $v_z$ from 0 to $v_F$ have equal weight.

In the case of purely diffusive transport, we use the $L$ diffusion equation for $\mu_L$ [23]. With a localized accumulation $\mu_L(z,t) \propto \delta(z)$ at time $t \approx 0$, the accumulation disperses according to the well-known solution

$$\mu_L(z,t) \propto \frac{1}{\sqrt{Dt}} \exp\left(-\frac{z^2}{4Dt}\right) \tag{11}$$

for $z > 0$ and $t > 0$ [ref]. Here, $D$ is the diffusion coefficient that equals $v_L^2 \tau/3$ in the case of a spherical **k**-space surface carrying the $L$ wavepackets, $v_L$ is their group velocity and $\tau$ their velocity relaxation time. To determine the current density, we apply Fick's law [23] $j_L = -D\partial_z \mu_L$ to Eq. (11) and obtain

$$r_z(t) \propto \Theta(t) \frac{z}{t} \frac{1}{\sqrt{Dt}} \exp\left(-\frac{z^2}{4Dt}\right). \tag{12}$$

***Ab-initio* estimate of the *L* velocity.** The orbital velocity is estimated for the bulk W in body-centered cubic (bcc) structure. The *ab-initio* self-consistent calculation of the electronic states is performed within the density functional theory by using the FLEUR code [59], which implements the full-potential linearly augmented plane wave (FLAPW) method [60]. The exchange correlation effect is included in the scheme of the generalized gradient approximation by using the Perdew-Burke-Ernzerhof functional [61][ref]. The lattice constant of the cubic unit cell is set to $5.96 a_0$, where $a_0$ is the Bohr radius. For the muffin-tin potential, we set $R_{\mathrm{MT}} = 2.5 a_0$ for the radius and $l_{\max} = 12$ for the maximum of the harmonic expansion. We set the plane-wave cutoffs for the interstitial region to $4.0 a_0^{-1}$, $10.1 a_0^{-1}$ and $12.2 a_0^{-1}$ for the basis set, the exchange-correlation functional and the charge density, respectively. For the **k**-points, a $16 \times 16 \times 16$ Monkhorst-Pack mesh is defined.

From the converged electronic structure, we obtain the maximally localized Wannier functions (MLWFs) by using the WANNIER90 code [62]. We use 18 Wannier states with $s, p_x, p_y, p_z, d_{z^2}, d_{x^2-y^2}, d_{xy}, d_{yz}, d_{zx}$ symmetries for spin up and down as the initial guess. The maximum of the inner (frozen) energy window is set 5 eV above the Fermi energy for the disentanglement, and the outer energy window is defined by the minimum and maximum energies of the 36 valence states obtained from the FLAPW calculation. The Hamiltonian, position, orbital and spin angular momenta operators are transformed from the FLAPW basis into the MLWF basis.

From this realistic tight-binding model, the orbital-momentum-weighted velocity averaged over the Fermi surface (FS) is calculated by

$$\langle v_\alpha L_\beta \rangle_{\mathrm{FS}} = \frac{\sum_{n\mathbf{k}} f'_{n\mathbf{k}} \langle n\mathbf{k}|(v_\alpha L_\beta + L_\beta v_\alpha)/2|n\mathbf{k}\rangle}{\sum_{n\mathbf{k}} f'_{n\mathbf{k}}}, \tag{13}$$

where $v_\alpha$ and $L_\beta$ are the $\alpha$ component of the velocity and the $\beta$ component of the orbital angular-momentum operators, respectively, $|n\mathbf{k}\rangle$ is the energy eigenstate of the Hamiltonian with band index $n$, and $f'_{n\mathbf{k}}$ is the energy derivative of the Fermi-Dirac distribution function. To polarize $L_\beta$, we add a small orbital Zeeman coupling along the $\beta$ direction to the bare Hamiltonian. We confirm that the result of Eq. (13) changes by less than 1% when the orbital Zeeman splitting is increased from 10 meV to 30 meV. The **k**-space integrals in Eq. (13) are performed on a $256 \times 256 \times 256$ mesh.

The orbital velocity is estimated by

$$\left\langle v_\alpha^{L_\beta} \right\rangle_{\mathrm{FS}} = \frac{\langle v_\alpha L_\beta \rangle_{\mathrm{FS}}}{\sqrt{\langle L_\beta^2 \rangle_{\mathrm{FS}}}}, \tag{14}$$

where $\langle L_\beta^2 \rangle_{\mathrm{FS}}$ is obtained by Eq. (13), but with $v_\alpha$ replaced by $L_\beta$. The result is shown in Fig. S11.

**Details of *ab-initio LC*C calculations.** A thin W stack of 19 bcc(110) atomic layers is calculated by the self-consistent *ab-initio* method using the same FLAPW parameters as for the bulk calculation above, except for the mesh of k-points for which we use a $24 \times 24$ Monkhorst-Pack mesh. For wannierization, we obtain 342

MLWFs, starting from Wannier states with $s, p_x, p_y, p_z, d_{z^2}, d_{x^2-y^2}, d_{xy}, d_{yz}, d_{zx}$ symmetries for spin up and down as the initial guess. We define the maximum of the inner window 2 eV above the Fermi energy.

From the Hamiltonian of the W thin film, the **k**-space orbital angular momentum texture at the Fermi surface is obtained by

$$\langle \mathbf{L}_{\text{top}} \rangle_{\text{FS}}(\mathbf{k}) = -4 k_B T \sum_n f'_{n\mathbf{k}} \langle \mathbf{L}_{\text{top}} \rangle_{n\mathbf{k}} \tag{15}$$

where $\langle \mathbf{L}_{\text{top}} \rangle_{n\mathbf{k}} = \langle n\mathbf{k} | \mathbf{L}_{\text{top}} | n\mathbf{k} \rangle$ is the expectation value of the orbital angular momentum for the two atoms on the top surface, $T = 300$ K is temperature, and $k_B$ is the Boltzmann constant.

To calculate the charge current due to $LC$C, we consider the orbital-dependent chemical potential

$$V = \frac{\mathcal{E}_{\beta\gamma}}{2}\left( r_\beta L_\gamma + L_\gamma r_\beta \right) \tag{16}$$

as a perturbation, where $r_\beta$ is the $\beta$ component of the position operator, which is well-defined along the perpendicular direction of the film, and $\mathcal{E}_{\beta\gamma}$ can be interpreted as an orbital-dependent electric field. The charge-current density along the $\alpha$ direction is given by the Kubo formula

$$\langle j_\alpha \rangle = -\frac{e}{V} \sum_{\mathbf{k}nn'} (f_{n\mathbf{k}} - f_{n'\mathbf{k}}) \operatorname{Re} \frac{\langle n\mathbf{k}|v_\alpha|n'\mathbf{k}\rangle \langle n'\mathbf{k}|V|n'\mathbf{k}\rangle}{E_{n\mathbf{k}} - E_{n'\mathbf{k}} + i\Gamma}, \tag{17}$$

where $V$ is the volume of the system, and $\Gamma = 25$ meV is a phenomenological broadening parameter for the energy spectrum. The $LC$C response is characterized by the tensor

$$\sigma^{L_\gamma}_{LCC,\alpha\beta} = -\frac{e}{2V} \sum_{\mathbf{k}nn'} (f_{n\mathbf{k}} - f_{n'\mathbf{k}}) \operatorname{Re} \frac{\langle n\mathbf{k}|v_\alpha|n'\mathbf{k}\rangle \langle n'\mathbf{k}|(r_\beta L_\gamma + L_\gamma r_\beta)|n'\mathbf{k}\rangle}{E_{n\mathbf{k}} - E_{n'\mathbf{k}} + i\Gamma}, \tag{18}$$

which relates $\langle j_\alpha \rangle$ and $\mathcal{E}_{\beta\gamma}$ by $\langle j_\alpha \rangle = \sigma^{L_\gamma}_{LCC,\alpha\beta} \mathcal{E}_{\beta\gamma}$. The **k**-space integral is performed on $400 \times 400$ mesh. The $z$-resolved $LC$C response is shown in Fig. S12.

**Acknowledgements**

We thank Giacomo Sala for fruitful discussions. TSS, RR and TK acknowledge funding by the German Research Foundation (DFG) through the collaborative research center SFB TRR 227 "Ultrafast spin dynamics" (project ID 328545488, projects A05 and B02) and financial support from the Horizon 2020 Framework Programme of the European Commission under FET-Open Grant No. 863155 (s-Nebula). FF and YM acknowledge the DFG collaborative research center SFB TRR 173/2 "Spin+X" (project ID 268565370, project A11). KA and HH acknowledge funding by JSPS (Grant Number 22H04964 and 20J20663) and Spintronics Research Network of Japan.

**References**

[1] Go, D., D. Jo, H.W. Lee, M. Klaui, and Y. Mokrousov *Orbitronics: Orbital currents in solids*. Epl, 2021. **135**: p. 37001.

[2] Miron, I.M., K. Garello, G. Gaudin, P.J. Zermatten, M.V. Costache, S. Auffret, S. Bandiera, B. Rodmacq, A. Schuhl, and P. Gambardella *Perpendicular switching of a single ferromagnetic layer induced by in-plane current injection*. Nature, 2011. **476**: p. 189.

[3] Ji, B., Y. Han, S. Liu, F. Tao, G. Zhang, Z. Fu, and C. Li *Several Key Technologies for 6G: Challenges and Opportunities*. IEEE Communications Standards Magazine, 2021. **5**: p. 44.

[4] Schwierz, F. and J.J. Liou *RF transistors: Recent developments and roadmap toward terahertz applications*. Solid-State Electronics, 2007. **51**: p. 1079.

[5] Vedmedenko, E.Y., R.K. Kawakami, D.D. Sheka, P. Gambardella, A. Kirilyuk, A. Hirohata, C. Binek, O. Chubykalo-Fesenko, S. Sanvito, B.J. Kirby, J. Grollier, K. Everschor-Sitte, T. Kampfrath, C.Y. You, and A. Berger *The 2020 magnetism roadmap*. Journal of Physics D-Applied Physics, 2020. **53**: p. 453001.

[6] Salemi, L., M. Berritta, A.K. Nandy, and P.M. Oppeneer *Orbitally dominated Rashba-Edelstein effect in noncentrosymmetric antiferromagnets*. Nat Commun, 2019. **10**: p. 5381.

[7] Johansson, A., B. Göbel, J. Henk, M. Bibes, and I. Mertig *Spin and orbital Edelstein effects in a two-dimensional electron gas: Theory and application to SrTiO 3 interfaces*. Physical Review Research, 2021. **3**: p. 013275.

[8] Bose, A., F. Kammerbauer, D. Go, Y. Mokrousov, G. Jakob, and M. Klaeui *Detection of long-range orbital-Hall torques*. arXiv preprint arXiv:2210.02283, 2022.

[9] Hayashi, H., D. Jo, D. Go, Y. Mokrousov, H.-W. Lee, and K. Ando *Observation of long-range orbital transport and giant orbital torque*. Communications Physics, 2023. **6**: p. 32.

[10] Zheng, Z.C., Q.X. Guo, D. Jo, D. Go, L.H. Wang, H.C. Chen, W. Yin, X.M. Wang, G.H. Yu, W. He, H.W. Lee, J. Teng, and T. Zhu *Magnetization switching driven by current-induced torque from weakly spin-orbit coupled Zr*. Physical Review Research, 2020. **2**: p. 013127.

[11] Kim, J., D. Go, H. Tsai, D. Jo, K. Kondou, H.-W. Lee, and Y. Otani *Nontrivial torque generation by orbital angular momentum injection in ferromagnetic-metal/Cu/Al 2 O 3 trilayers*. Physical Review B, 2021. **103**: p. L020407.

[12] Lee, D., D. Go, H.J. Park, W. Jeong, H.W. Ko, D. Yun, D. Jo, S. Lee, G. Go, J.H. Oh, K.J. Kim, B.G. Park, B.C. Min, H.C. Koo, H.W. Lee, O. Lee, and K.J. Lee *Orbital torque in magnetic bilayers*. Nat Commun, 2021. **12**: p. 6710.

[13] Choi, Y.-G., D. Jo, K.-H. Ko, D. Go, K.-H. Kim, H.G. Park, C. Kim, B.-C. Min, G.-M. Choi, and H.-W. Lee *Observation of the orbital Hall effect in a light metal Ti*. arXiv preprint arXiv:2109.14847, 2021.

[14] Go, D., D. Jo, K.-W. Kim, S. Lee, M.-G. Kang, B.-G. Park, S. Blügel, H.-W. Lee, and Y. Mokrousov *Long-Range Orbital Transport in Ferromagnets*. arXiv preprint arXiv:2106.07928, 2021.

[15] An, H., Y. Kageyama, Y. Kanno, N. Enishi, and K. Ando *Spin-torque generator engineered by natural oxidation of Cu*. Nat Commun, 2016. **7**: p. 13069.

[16] Go, D., D. Jo, C. Kim, and H.W. Lee *Intrinsic Spin and Orbital Hall Effects from Orbital Texture*. Phys Rev Lett, 2018. **121**: p. 086602.

[17] Go, D., F. Freimuth, J.P. Hanke, F. Xue, O. Gomonay, K.J. Lee, S. Blugel, P.M. Haney, H.W. Lee, and Y. Mokrousov *Theory of Current-Induced Angular Momentum Transfer Dynamics in Spin-Orbit Coupled Systems*. Phys Rev Res, 2020. **2**: p. 033401.

[18] Go, D. and H.W. Lee *Orbital torque: Torque generation by orbital current injection*. Physical Review Research, 2020. **2**: p. 013177.

[19] Tazaki, Y., Y. Kageyama, H. Hayashi, T. Harumoto, T. Gao, J. Shi, and K. Ando *Current-induced torque originating from orbital current*. arXiv preprint arXiv:2004.09165, 2020.

[20] Lee, S., M.G. Kang, D. Go, D. Kim, J.H. Kang, T. Lee, G.H. Lee, J. Kang, N.J. Lee, Y. Mokrousov, S. Kim, K.J. Kim, K.J. Lee, and B.G. Park *Efficient conversion of orbital Hall current to spin current for spin-orbit torque switching*. Communications Physics, 2021. **4**: p. 1.

[21] Liao, L., F. Xue, L. Han, J. Kim, R. Zhang, L. Li, J. Liu, X. Kou, C. Song, and F. Pan *Efficient orbital torque in polycrystalline ferromagnetic– metal/Ru/Al 2 O 3 stacks: Theory and experiment*. Physical Review B, 2022. **105**: p. 104434.

[22] Xiao, Z.Y., Y.J. Li, W. Zhang, Y.J. Han, D. Li, Q. Chen, Z.M. Zeng, Z.Y. Quan, and X.H. Xu *Enhancement of torque efficiency and spin Hall angle driven collaboratively by orbital torque and spin-orbit torque*. Applied Physics Letters, 2022. **121**: p. 072404.
[23] Sala, G. and P. Gambardella *Giant orbital Hall effect and orbital-to-spin conversion in 3 d, 5 d, and 4 f metallic heterostructures*. Physical Review Research, 2022. **4**: p. 033037.
[24] Ding, S., A. Ross, D. Go, L. Baldrati, Z. Ren, F. Freimuth, S. Becker, F. Kammerbauer, J. Yang, G. Jakob, Y. Mokrousov, and M. Klaui *Harnessing Orbital-to-Spin Conversion of Interfacial Orbital Currents for Efficient Spin-Orbit Torques*. Phys Rev Lett, 2020. **125**: p. 177201.
[25] Manchon, A., J. Zelezny, I.M. Miron, T. Jungwirth, J. Sinova, A. Thiaville, K. Garello, and P. Gambardella *Current-induced spin-orbit torques in ferromagnetic and antiferromagnetic systems*. Reviews of Modern Physics, 2019. **91**: p. 035004.
[26] Seifert, T.S., L. Cheng, Z. Wei, T. Kampfrath, and J. Qi, *Spintronic sources of ultrashort terahertz electromagnetic pulses*. 2022, AIP Publishing LLC. p. 180401.
[27] Xu, Y., F. Zhang, Y. Liu, R. Xu, Y. Jiang, H. Cheng, A. Fert, and W. Zhao *Inverse Orbital Hall Effect Discovered from Light-Induced Terahertz Emission*. arXiv preprint arXiv:2208.01866, 2022.
[28] Seifert, T., S. Jaiswal, U. Martens, J. Hannegan, L. Braun, P. Maldonado, F. Freimuth, A. Kronenberg, J. Henrizi, I. Radu, E. Beaurepaire, Y. Mokrousov, P.M. Oppeneer, M. Jourdan, G. Jakob, D. Turchinovich, L.M. Hayden, M. Wolf, M. Munzenberg, M. Klaui, and T. Kampfrath *Efficient metallic spintronic emitters of ultrabroadband terahertz radiation*. Nature Photonics, 2016. **10**: p. 483.
[29] Kampfrath, T., M. Battiato, P. Maldonado, G. Eilers, J. Notzold, S. Mahrlein, V. Zbarsky, F. Freimuth, Y. Mokrousov, S. Blugel, M. Wolf, I. Radu, P.M. Oppeneer, and M. Munzenberg *Terahertz spin current pulses controlled by magnetic heterostructures*. Nat Nanotechnol, 2013. **8**: p. 256.
[30] Jungfleisch, M.B., Q. Zhang, W. Zhang, J.E. Pearson, R.D. Schaller, H. Wen, and A. Hoffmann *Control of Terahertz Emission by Ultrafast Spin-Charge Current Conversion at Rashba Interfaces*. Phys Rev Lett, 2018. **120**: p. 207207.
[31] Zhou, C., Y.P. Liu, Z. Wang, S.J. Ma, M.W. Jia, R.Q. Wu, L. Zhou, W. Zhang, M.K. Liu, Y.Z. Wu, and J. Qi *Broadband Terahertz Generation via the Interface Inverse Rashba-Edelstein Effect*. Phys Rev Lett, 2018. **121**: p. 086801.
[32] Rouzegar, R., L. Brandt, L. Nadvornik, D.A. Reiss, A.L. Chekhov, O. Gueckstock, C. In, M. Wolf, T.S. Seifert, P.W. Brouwer, G. Woltersdorf, and T. Kampfrath *Laser-induced terahertz spin transport in magnetic nanostructures arises from the same force as ultrafast demagnetization*. Physical Review B, 2022. **106**: p. 144427.
[33] Lichtenberg, T., M. Beens, M.H. Jansen, B. Koopmans, and R.A. Duine *Probing optically induced spin currents using terahertz spin waves in noncollinear magnetic bilayers*. Physical Review B, 2022. **105**: p. 144416.
[34] Mueller, B.Y. and B. Rethfeld *Thermodynamic µT model of ultrafast magnetization dynamics*. Physical Review B, 2014. **90**: p. 144420.
[35] Choi, G.M., B.C. Min, K.J. Lee, and D.G. Cahill *Spin current generated by thermally driven ultrafast demagnetization*. Nat Commun, 2014. **5**: p. 4334.
[36] Boeglin, C., E. Beaurepaire, V. Halte, V. Lopez-Flores, C. Stamm, N. Pontius, H.A. Durr, and J.Y. Bigot *Distinguishing the ultrafast dynamics of spin and orbital moments in solids*. Nature, 2010. **465**: p. 458.
[37] Stamm, C., N. Pontius, T. Kachel, M. Wietstruk, and H.A. Durr *Femtosecond x-ray absorption spectroscopy of spin and orbital angular momentum in photoexcited Ni films during ultrafast demagnetization*. Physical Review B, 2010. **81**: p. 104425.
[38] Hennecke, M., I. Radu, R. Abrudan, T. Kachel, K. Holldack, R. Mitzner, A. Tsukamoto, and S. Eisebitt *Angular Momentum Flow During Ultrafast Demagnetization of a Ferrimagnet*. Phys Rev Lett, 2019. **122**: p. 157202.
[39] Zhang, W., P. Maldonado, Z. Jin, T.S. Seifert, J. Arabski, G. Schmerber, E. Beaurepaire, M. Bonn, T. Kampfrath, P.M. Oppeneer, and D. Turchinovich *Ultrafast terahertz magnetometry*. Nat Commun, 2020. **11**: p. 4247.
[40] Salemi, L. and P.M. Oppeneer *First-principles theory of intrinsic spin and orbital Hall and Nernst effects in metallic monoatomic crystals*. Physical Review Materials, 2022. **6**: p. 095001.

[41] Seifert, T.S., N.M. Tran, O. Gueckstock, S.M. Rouzegar, L. Nadvornik, S. Jaiswal, G. Jakob, V.V. Temnov, M. Münzenberg, M. Wolf, M. Kläui, and T. Kampfrath *Terahertz spectroscopy for all-optical spintronic characterization of the spin-Hall-effect metals Pt, W and Cu80Ir20*. Journal of Physics D: Applied Physics, 2018. **51**: p. 364003.

[42] Huber, R., A. Brodschelm, F. Tauser, and A. Leitenstorfer *Generation and field-resolved detection of femtosecond electromagnetic pulses tunable up to 41 THz*. Applied Physics Letters, 2000. **76**: p. 3191.

[43] Braun, L., G. Mussler, A. Hruban, M. Konczykowski, T. Schumann, M. Wolf, M. Münzenberg, L. Perfetti, and T. Kampfrath *Ultrafast photocurrents at the surface of the three-dimensional topological insulator $Bi_2Se_3$*. Nature communications, 2016. **7**: p. 1.

[44] Gueckstock, O., L. Nadvornik, M. Gradhand, **T.S. Seifert**, G. Bierhance, R. Rouzegar, M. Wolf, M. Vafaee, J. Cramer, M.A. Syskaki, G. Woltersdorf, I. Mertig, G. Jakob, M. Klaui, and T. Kampfrath *Terahertz Spin-to-Charge Conversion by Interfacial Skew Scattering in Metallic Bilayers*. Adv Mater, 2021. **33**: p. e2006281.

[45] Tanaka, T., H. Kontani, M. Naito, T. Naito, D.S. Hirashima, K. Yamada, and J. Inoue *Intrinsic spin Hall effect and orbital Hall effect in 4 d and 5 d transition metals*. Physical Review B, 2008. **77**: p. 165117.

[46] Demasius, K.-U., T. Phung, W. Zhang, B.P. Hughes, S.-H. Yang, A. Kellock, W. Han, A. Pushp, and S.S. Parkin *Enhanced spin–orbit torques by oxygen incorporation in tungsten films*. Nature communications, 2016. **7**: p. 10644.

[47] Ding, S., Z. Liang, D. Go, C. Yun, M. Xue, Z. Liu, S. Becker, W. Yang, H. Du, C. Wang, Y. Yang, G. Jakob, M. Klaui, Y. Mokrousov, and J. Yang *Observation of the Orbital Rashba-Edelstein Magnetoresistance*. Phys Rev Lett, 2022. **128**: p. 067201.

[48] Santos, E., J. Abrão, D. Go, L. de Assis, Y. Mokrousov, J. Mendes, and A. Azevedo *Inverse Orbital Torque via Spin-Orbital Entangled States*. arXiv preprint arXiv:2204.01825, 2022.

[49] Okano, G., M. Matsuo, Y. Ohnuma, S. Maekawa, and Y. Nozaki *Nonreciprocal Spin Current Generation in Surface-Oxidized Copper Films*. Phys Rev Lett, 2019. **122**: p. 217701.

[50] Yoda, T., T. Yokoyama, and S. Murakami *Orbital Edelstein Effect as a Condensed-Matter Analog of Solenoids*. Nano Letters, 2018. **18**: p. 916.

[51] Go, D., J.P. Hanke, P.M. Buhl, F. Freimuth, G. Bihlmayer, H.W. Lee, Y. Mokrousov, and S. Blugel *Toward surface orbitronics: giant orbital magnetism from the orbital Rashba effect at the surface of sp-metals*. Scientific Reports, 2017. **7**: p. 1.

[52] Hixson, R. and M. Winkler *Thermophysical properties of solid and liquid tungsten*. International Journal of Thermophysics, 1990. **11**: p. 709.

[53] Wu, Y., M. Elyasi, X. Qiu, M. Chen, Y. Liu, L. Ke, and H. Yang *High-Performance THz Emitters Based on Ferromagnetic/Nonmagnetic Heterostructures*. Adv Mater, 2017. **29**: p. 1603031.

[54] Zhang, S., Z. Jin, Z. Zhu, W. Zhu, Z. Zhang, G. Ma, and J. Yao *Bursts of efficient terahertz radiation with saturation effect from metal-based ferromagnetic heterostructures*. Journal of Physics D: Applied Physics, 2017. **51**: p. 034001.

[55] Seifert, T.S., S. Jaiswal, J. Barker, S.T. Weber, I. Razdolski, J. Cramer, O. Gueckstock, S.F. Maehrlein, L. Nadvornik, S. Watanabe, C. Ciccarelli, A. Melnikov, G. Jakob, M. Munzenberg, S.T.B. Goennenwein, G. Woltersdorf, B. Rethfeld, P.W. Brouwer, M. Wolf, M. Klaui, and T. Kampfrath *Femtosecond formation dynamics of the spin Seebeck effect revealed by terahertz spectroscopy*. Nat Commun, 2018. **9**: p. 2899.

[56] Naftaly, M. and R.E. Miles *Terahertz time-domain spectroscopy of silicate glasses and the relationship to material properties*. Journal of Applied Physics, 2007. **102**: p. 043517.

[57] Lin, Z., L.V. Zhigilei, and V. Celli *Electron-phonon coupling and electron heat capacity of metals under conditions of strong electron-phonon nonequilibrium*. Physical Review B, 2008. **77**: p. 075133.

[58] Ordal, M.A., L.L. Long, R.J. Bell, S.E. Bell, R.R. Bell, R.W. Alexander, Jr., and C.A. Ward *Optical properties of the metals Al, Co, Cu, Au, Fe, Pb, Ni, Pd, Pt, Ag, Ti, and W in the infrared and far infrared*. Appl Opt, 1983. **22**: p. 1099.

[59] Available from: https://www.flapw.de.

[60] Wimmer, E., H. Krakauer, M. Weinert, and A. Freeman *Full-potential self-consistent linearized-augmented-plane-wave method for calculating the electronic structure of molecules and surfaces: $O_2$ molecule*. Physical Review B, 1981. **24**: p. 864.

[61] Perdew, J.P., K. Burke, and M. Ernzerhof *Generalized gradient approximation made simple*. Physical review letters, 1996. **77**: p. 3865.

[62] Pizzi, G., V. Vitale, R. Arita, S. Blügel, F. Freimuth, G. Géranton, M. Gibertini, D. Gresch, C. Johnson, and T. Koretsune *Wannier90 as a community code: new features and applications*. Journal of Physics: Condensed Matter, 2020. **32**: p. 165902.

[63] Nádvorník, L., M. Borchert, L. Brandt, R. Schlitz, K.A. de Mare, K. Výborný, I. Mertig, G. Jakob, M. Kläui, and S.T. Goennenwein *Broadband terahertz probes of anisotropic magnetoresistance disentangle extrinsic and intrinsic contributions*. Physical Review X, 2021. **11**: p. 021030.

[64] Zak, J., E. Moog, C. Liu, and S. Bader *Universal approach to magneto-optics*. Journal of Magnetism and Magnetic Materials, 1990. **89**: p. 107.

**Supplementary Materials**

First, we summarize briefly the content of the Supplementary Materials before showing the corresponding data:

- Samples on Si show qualitatively the same THz emission waveforms for Ni with Pt, W and Ti. Most importantly, the strong change in W dynamics is also observed on Si (Fig. S1). However, the THz waveforms of Si vs glass differ in the details, which might be related to slightly changed transport times.
- Drude scattering times are estimated to be $\ll 50$ fs for all studied samples (Figs. S2). None of the samples showed any indication of a drastically different Drude scattering time compared to all other samples.
- Emitted THz signals are found to be linearly polarized and perpendicular to the sample magnetization (Fig. S3).
- Pump-polarization dependent studies (pump helicity and linear polarization direction) show a minor impact on the measured THz emission signal (Figs. S4).
- We perform THz emission measurements upon reversing the sample. Only the pure Ni film shows a dominant contribution even in sample rotation, which we ascribe to SIA or magnetic dipole radiation (Fig. S5) [32, 39].
- For all Py-based bilayer samples, we find almost identical THz emission waveform shapes even for PM thicknesses of 20 nm (Fig. S6).
- For Ni|Ti samples, we find almost identical THz emission dynamics to Ni|Pt (Fig. S7).
- *Ab-initio* claculations of the orbital polarization close to the W-layer surface, the estimated velocity of the orbital current and the interface concentrated $LC$C in a W film are shown in Figs. S8-S10.
- A comparison of purely ballistic vs purely diffusive motion for orbitally polarized wave packets is shown in Fig. S11.
- Currents driven by pump light gradients in thick films of Ni|W and Ni|Ti can be neglected (Fig. S12).
- All fluence dependencies are to a good approximation linear (Fig S13) with minor sublinearities overserved for Ni|Ti and Ni|W samples. Related to that, only minor changes in the THz waveform dynamics can be observed for different pumping fluences (Fig S13)
- Cu has only minor impact on the emitted THz waveforms (Fig. S14), either as a spacer layer or as a capping layer as confirmed by comparison to the same sample without Cu.
- All data in the Supplementary Materials was measured with a 1 mm thick ZnTe(110) detection crystal.

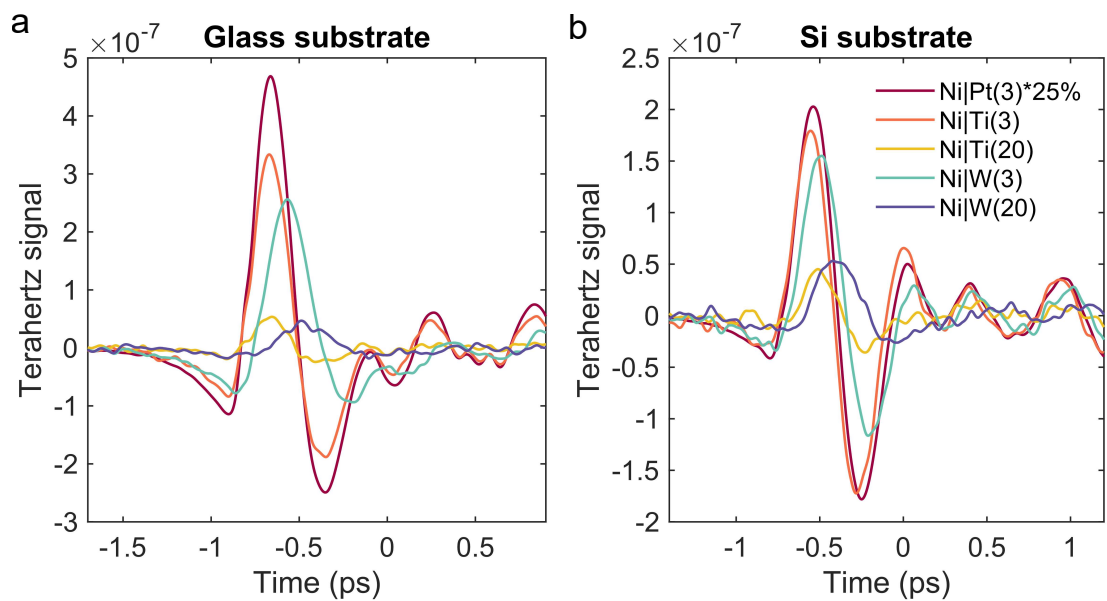

**FIGURE S1: Si vs glass substrate. a,** Terahertz-emission waveforms from Ni|PM stacks on Si substrates. THz waveforms for Si based samples are multiplied by -1 to account for the reversed sample orientation due to the intransparency of the Si substrate for the pump pulse. **b,** Terahertz-emission waveforms from Ni|PM stacks on glass substrates. Film thicknesses in nanometers are given as numerals in parenthesis, except for Ni layers, which are always 5 nm thick. Note the rescaling of the Ni|Pt sample THz waveforms.

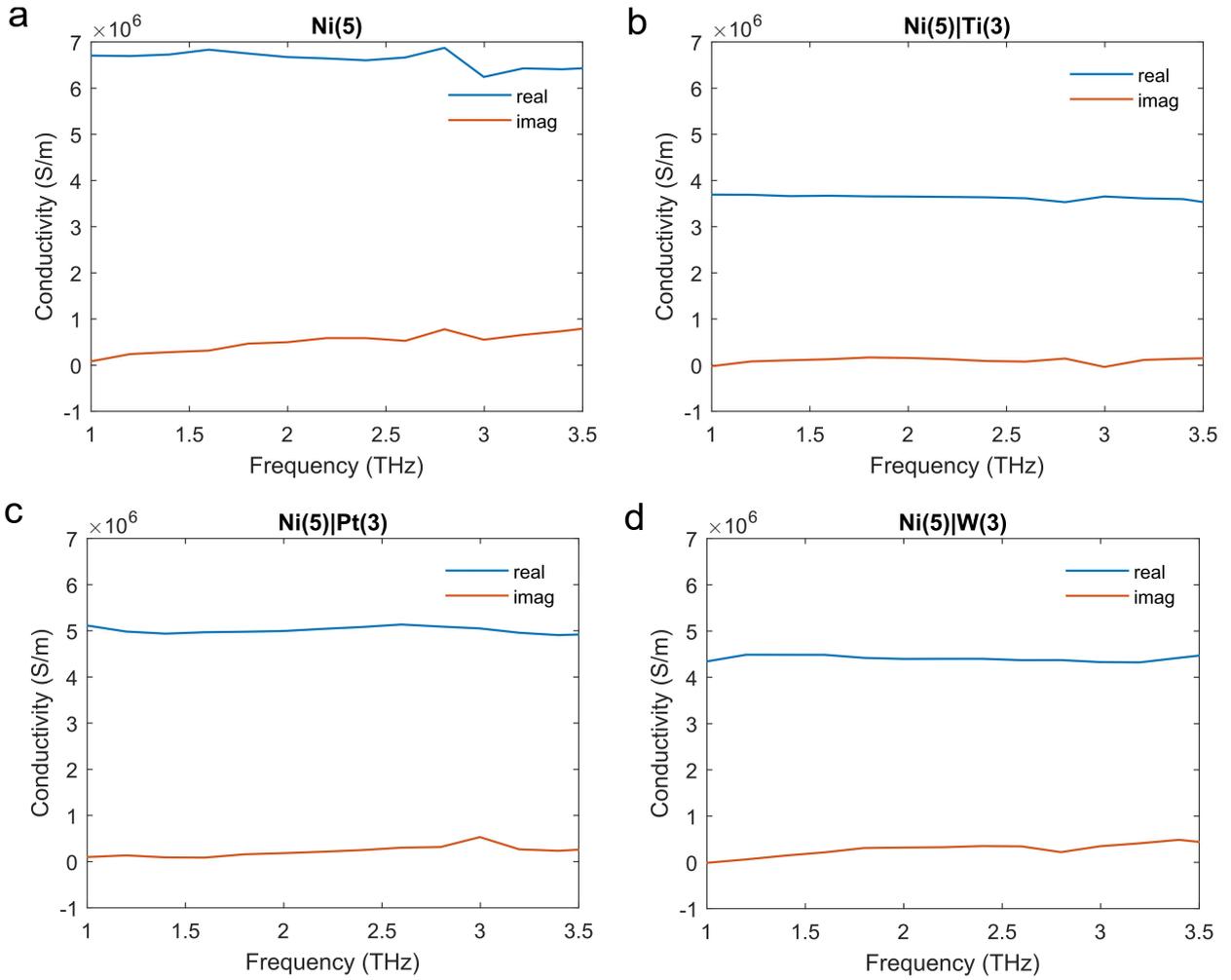

**FIGURE S2: Terahertz conductivities for samples on glass.** Mean complex-valued terahertz conductivities obtained from terahertz transmission measurements for **a**, Ni, **b**, Ni|Ti, **c**, Ni|Pt and **d**, Ni|W samples. For the extraction, a thin-film formula is applied [41] and a terahertz refractive index of 2.1 for glass is assumed. Film thicknesses in nanometers are given as numerals in parentheses. In all panels, the extrapolated real (blue solid line) and imaginary parts (red) of the conductivity are expected to cross at frequencies $\Gamma/2\pi \gg 3$ THz. For a Drude-like conductivity, it follows that the current relaxation time $1/\Gamma$ is $\ll 50$ fs [63].

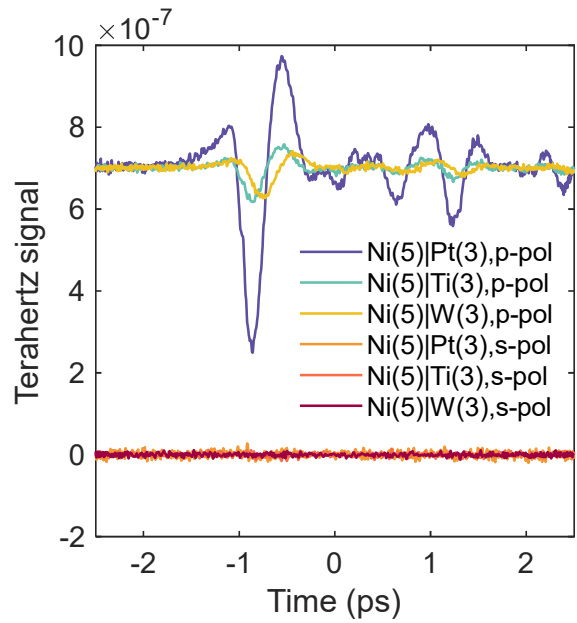

**FIGURE S3: Polarization state of the THz emission signal.** Samples are magnetized along the s-direction and pump pulses are polarized along the p-direction. Film thicknesses in nanometers are given as numerals in parenthesis. The vertically offset upper curves show the p-polarized THz emission signal, whereas the lower curves show the s-polarized THz emission signal. This figure implies that the THz emission signal is primarily polarized linearly and perpendicular to the sample magnetization.

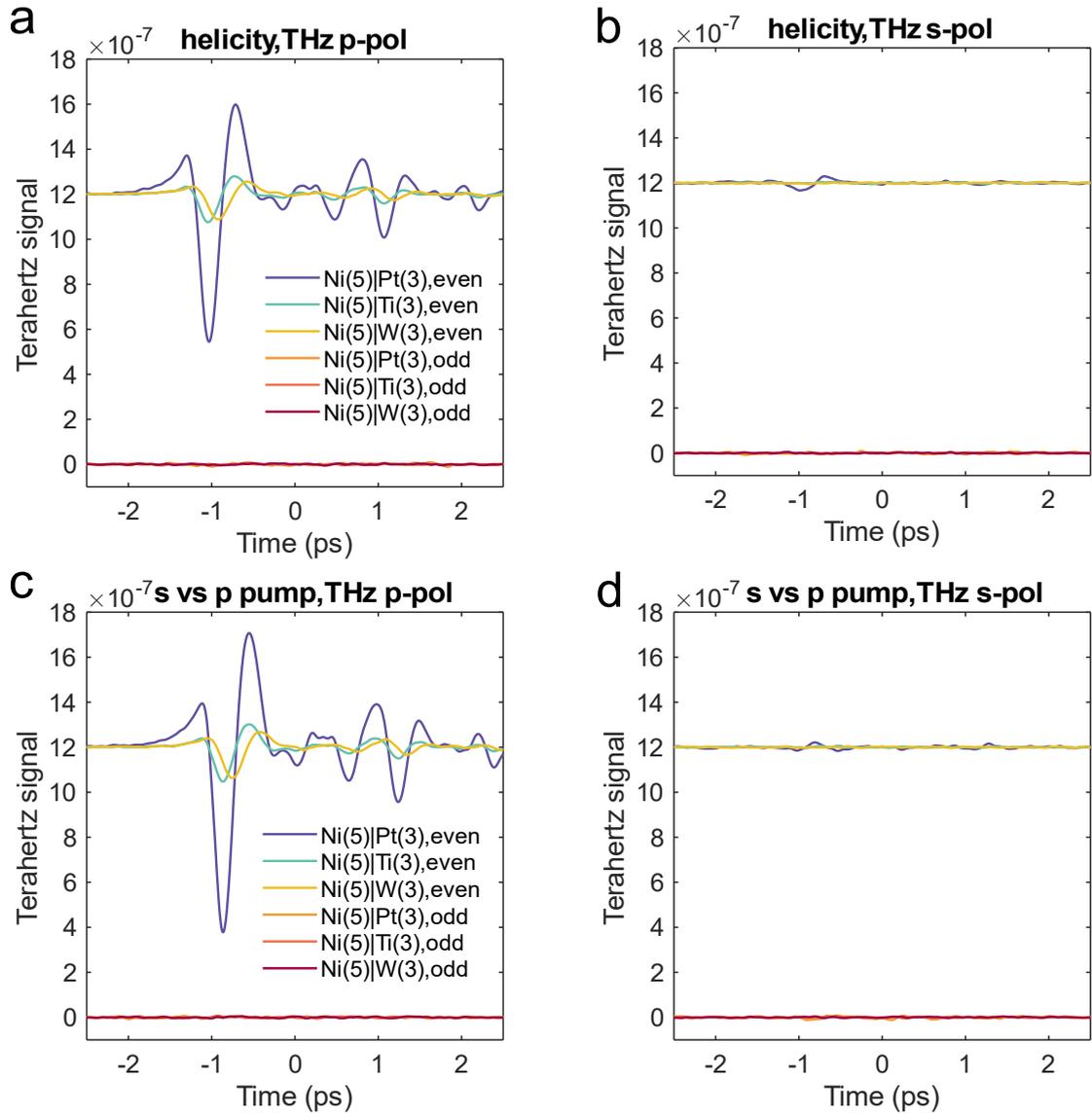

**FIGURE S4: Impact of pump polarization. a**, **b**, Circular pump polarization. Terahertz emission signals even and odd with respect to pump helicity (right- and left-circular) for terahertz emission polarized along the p-direction (panel a) and s-direction (panel b). **c**, **d**, Linear pump polarization. Terahertz emission signals even and odd with respect to linear pump polarization (s- and p-polarized) for terahertz emission polarized along the p-direction (panel c) and s-direction (panel d). Samples are magnetized along the s-direction. Film thicknesses in nanometers are given as numerals in parenthesis. This figure implies that the THz emission signal does neither depend on the pump helicity (left- vs right-circular; panels a, b) nor on the linear polarization direction (s vs p; panels c, d).

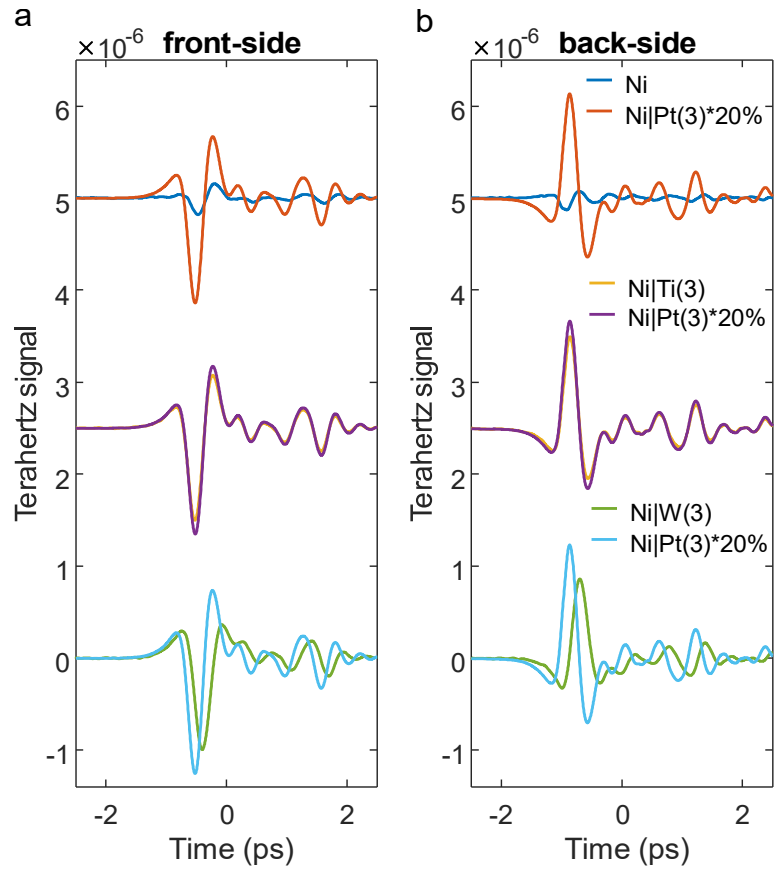

**FIGURE S5: Front-side vs back-side pump geometry. a**, Samples pumped from the front side. **b**, Samples pumped from the back side. The back-side pumping is defined as the direction where the pump pulse first traverses the substrate before exciting the sample and is the standard direction used for all measurements throughout this work. Film thicknesses in nanometers are given as numerals in parenthesis, except for the Ni layers, which are always 5 nm thick.

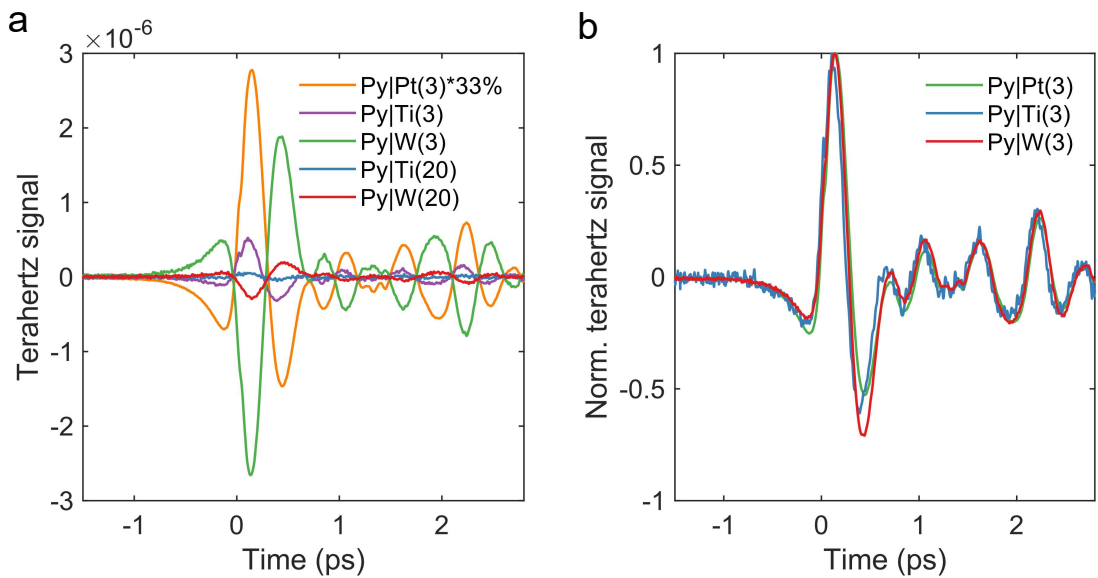

**FIGURE S6: Terahertz emission signals for Py-based samples. a**, Terahertz emission signals from thicker Ti and W layers on Py. **b**, Normalized THz emission signals from the data shown in Fig. 2a in the main text. Film thicknesses in nanometers are given as numerals in parentheses, except for the Py layers, which are always 5 nm thick.

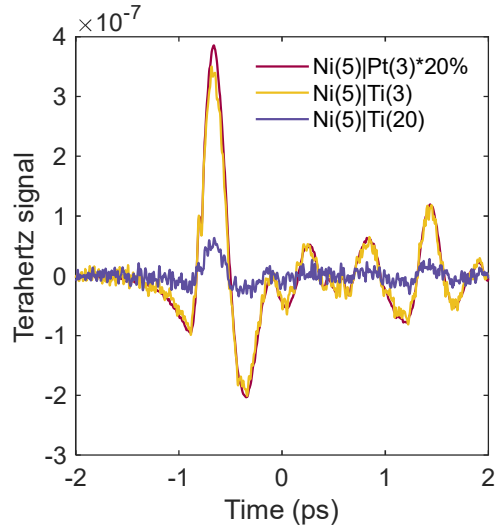

**FIGURE S7: Ni|Pt vs Ni|Ti.** Film thicknesses in nanometers are given as numerals in parentheses. Note the rescaling of the Ni|Pt sample waveform.

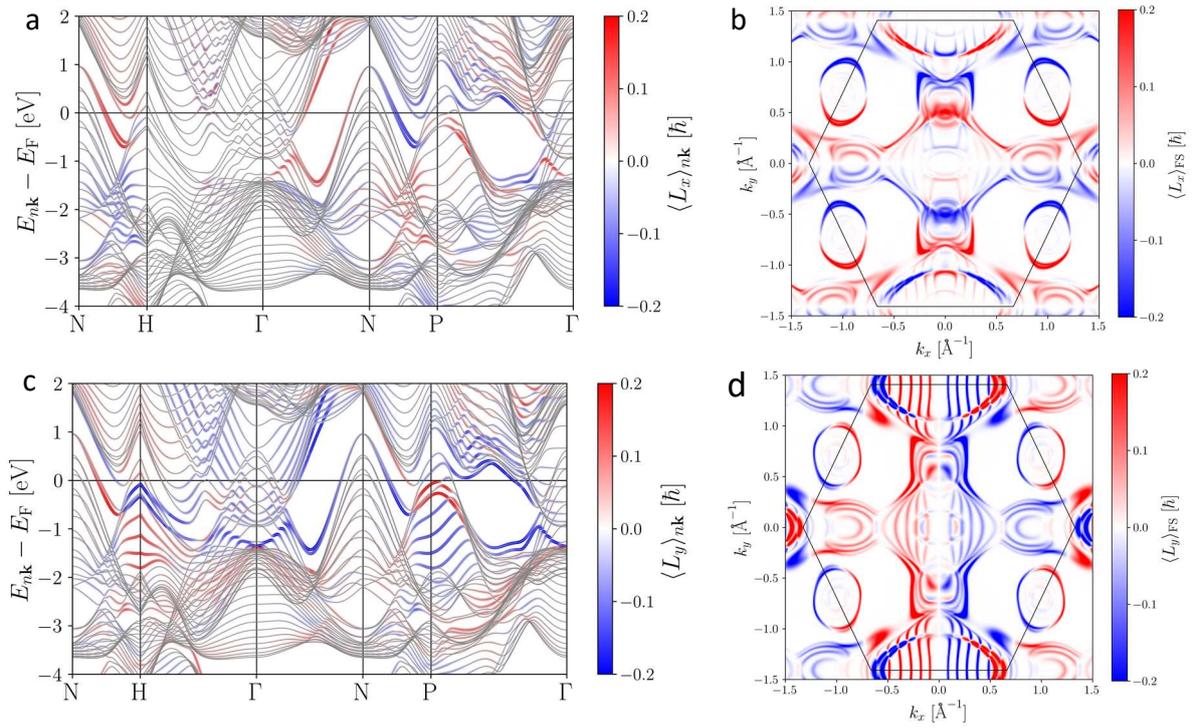

**FIGURE S8: Ab-initio calculation of the electronic structure and L texture at the surface of a W thin film.** The film consists of 19 layers of W atoms in bcc(110) stacking, and **L** is evaluated for the two topmost surface atoms. **a**, Electronic band structure $E_{n\mathbf{k}}$ (grey lines) together with the expectation value of $L_x$ for individual states. **b**, **k**-space texture of $L_x$, **c**, $L_y$ and **d**, $L_z$.

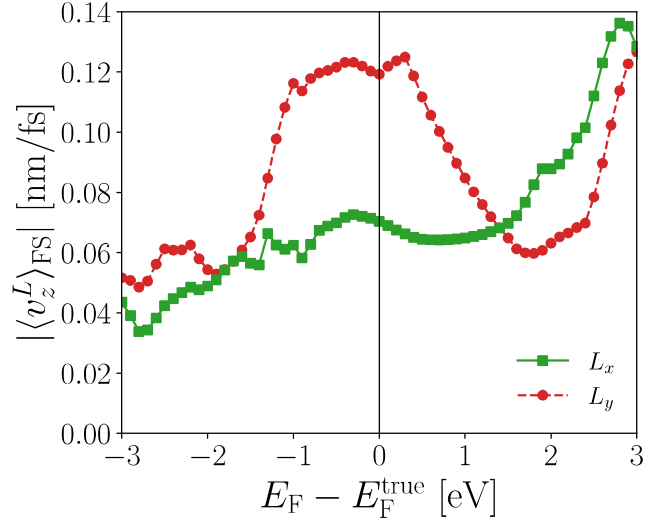

**FIGURE S9: Orbital velocity at the Fermi surface.** The orbital velocity is calculated for states at the Fermi surface for bulk W in bcc structure for different values of the Fermi energy with respect to the true Fermi energy (see Methods). The coordinate is defined such that $x \parallel [001]$, $y \parallel [1\bar{1}0]$, $z \parallel [110]$.

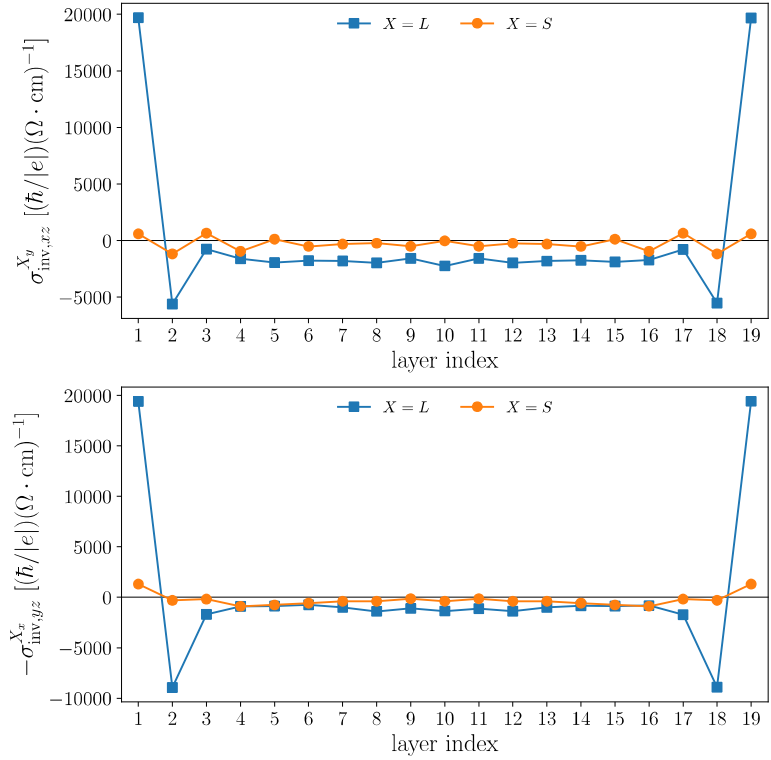

**FIGURE S10: Ab-initio calculation of the charge current response to an orbital-dependent chemical potential in a W thin film.** The film consists of 19 layers of W atoms in bcc(110) stacking. The response is calculated by the Kubo formula (see Methods). The charge-current response is strongly pronounced at the W/vacuum surfaces and has positive sign, which coincides with the sign of the inverse spin Hall effect of Pt. The coordinate is defined such that $x \parallel [001]$, $y \parallel [1\bar{1}0]$, $z \parallel [110]$.

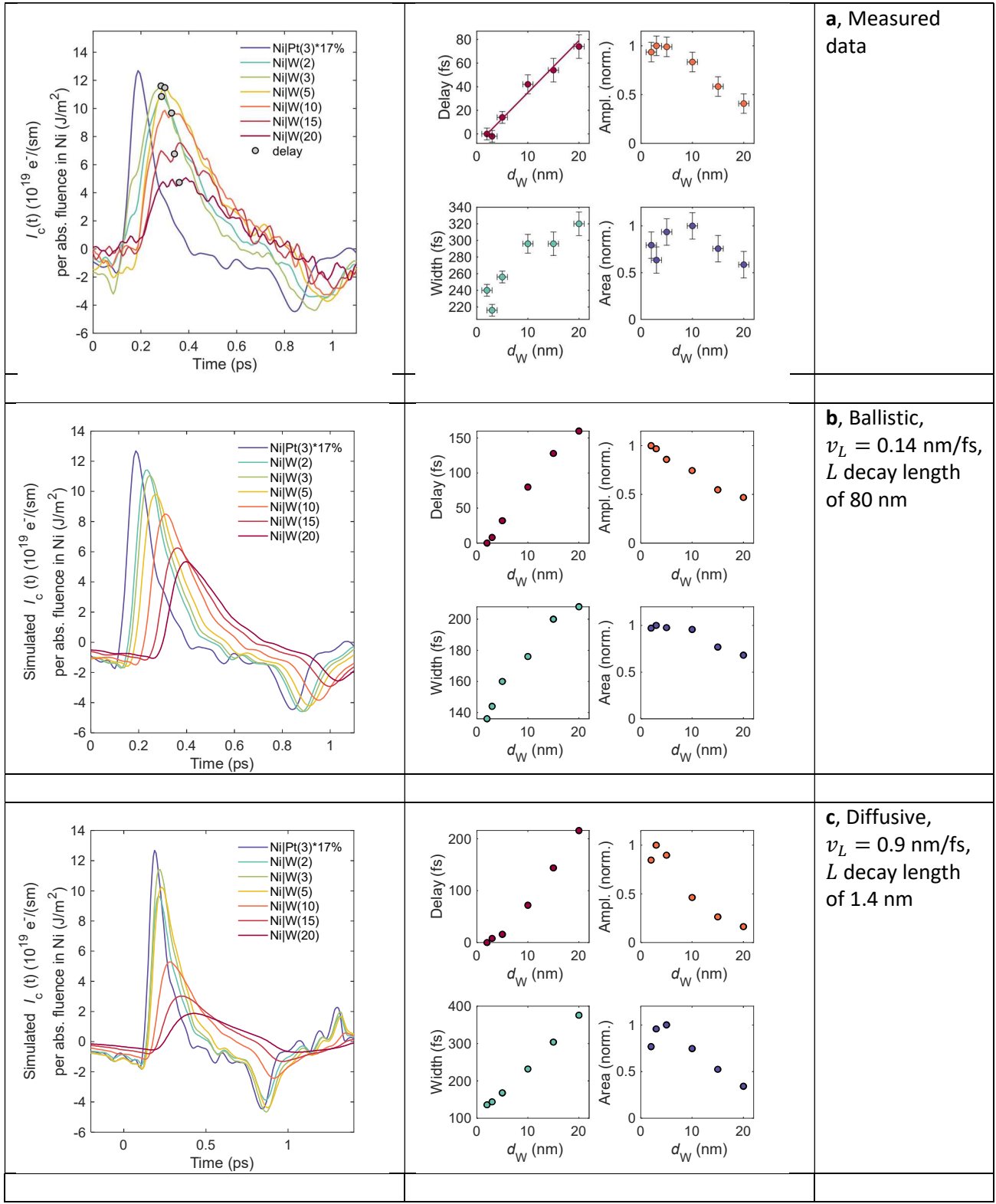

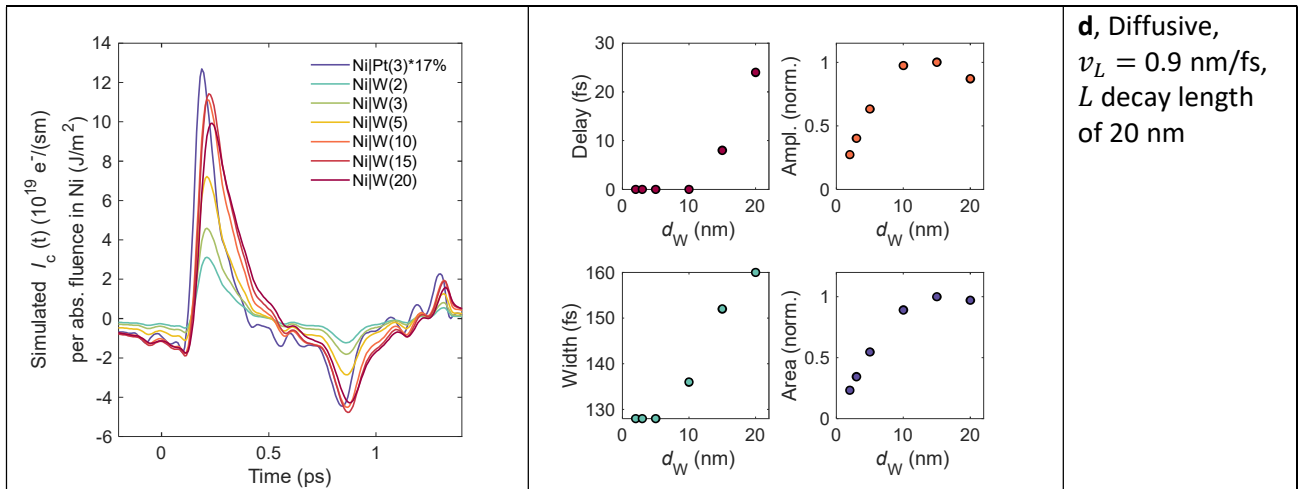

**FIGURE S11: Comparison of modeled charge currents for different propagation regimes. a,** Experimental data. **b,** Ballistic limit. **c** and **d,** Diffusive limit. The right-hand side of each panel shows the extracted charge current parameters as defined in Fig. 4 of the main text.

**d,** Diffusive, $v_L = 0.9$ nm/fs, $L$ decay length of 20 nm

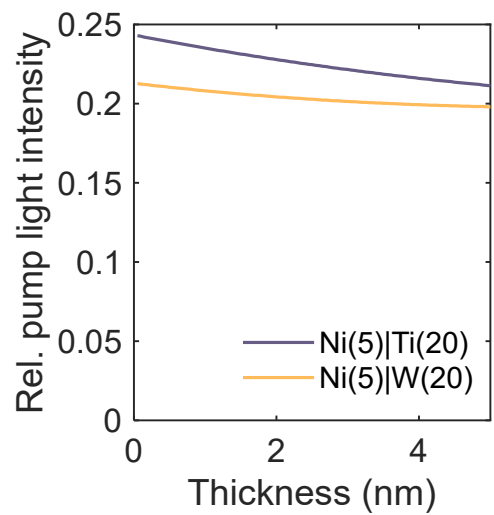

**FIGURE S12:** Calculated pump-intensity gradient in Ni for Ni(5)|Ti(20) and Ni(5)|W(20) samples, which are the thickest samples measured. However, even in these thickest samples, the pump-light gradient is minor. The calculation is based on a transfer matrix formalism [64]. Film thicknesses in nanometers are given as numerals in parentheses.

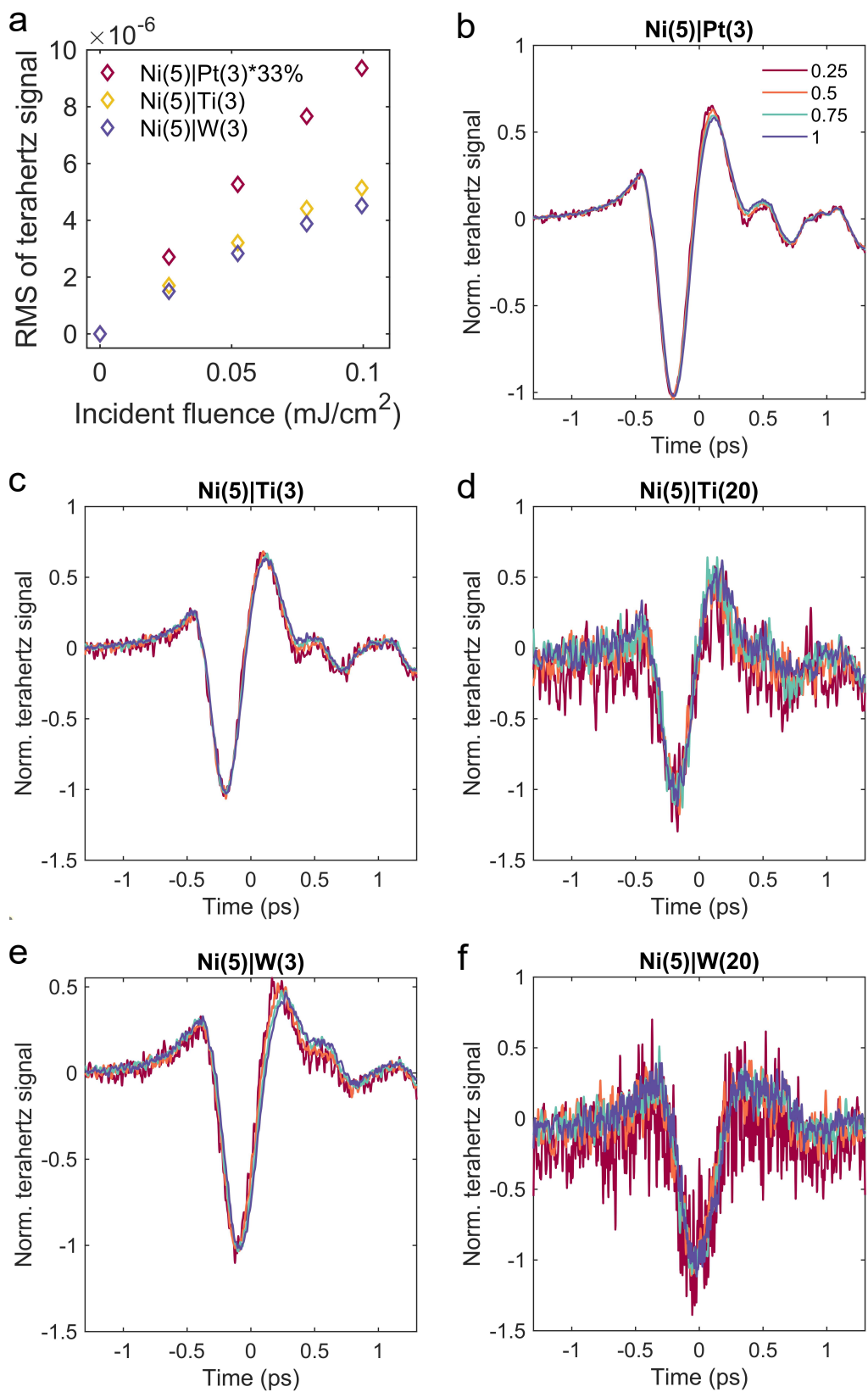

**FIGURE S13: Pump fluence dependencies. a**, Fluence dependencies of Ni capped with Pt, W or Ti. The data was contracted by taking the root mean square (RMS) of the time-domain traces. **b-f**, Normalized THz emission signals for different pump fluences. The different colors correspond to the fluence levels applied (25%: red, 50%: orange, 75%: cyan and 100%: blue). Film thicknesses in nanometers are given as numerals in parentheses.

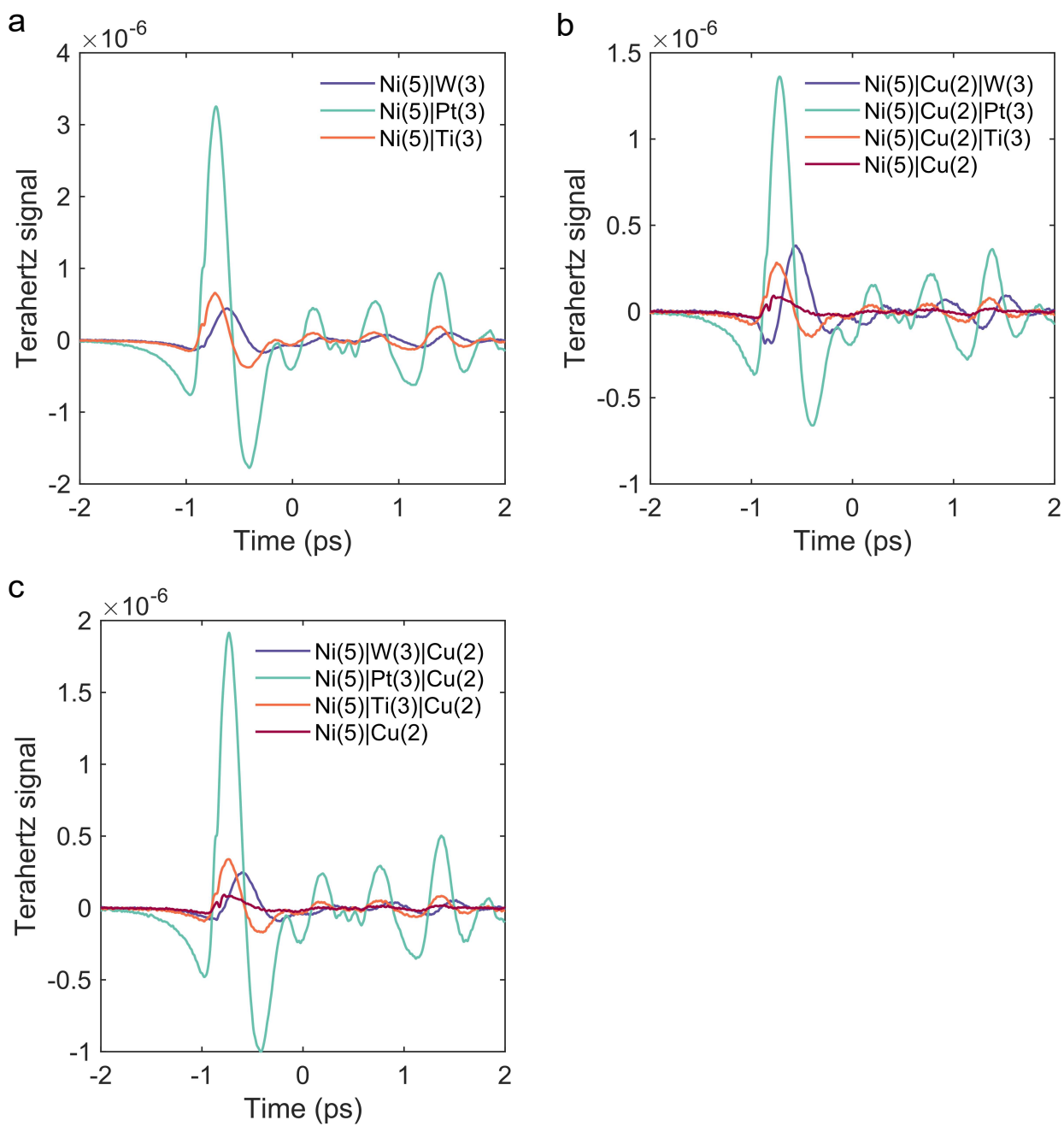

**FIGURE S14: Impact of Cu interlayers and capping layers. a**, Reference samples without Cu. **b**, Samples with Cu intermediate layer. **c**, Samples with Cu capping layer. Film thicknesses in nanometers are given as numerals in parentheses.

| Sample | Absorptance | Absorbed fluence in the FM layer (mJ/cm$^2$) | Absorbed fluence in the PM layer (mJ/cm$^2$) | Conductivity (MS/m) |
|---|---|---|---|---|
| Glass\| Ti(50) | - | - | - | 1.6 |
| Glass\| Ni(5)\|W(20) | 0.52 | 0.06 | 0.20 | 5.1 |
| Glass\| Ni(5)\|Pt(3) | 0.63 | 0.25 | 0.06 | 3.6 |
| Glass\| Ni(5)\|Ti(3) | 0.58 | 0.19 | 0.10 | 2.2 |
| Glass\| Ni(5)\|W(3) | 0.58 | 0.19 | 0.10 | 2.1 |
| Glass\| Ni(5)\|Ti(20) | 0.51 | 0.06 | 0.20 | 1.6 |
| Glass\| Ni(5) | 0.51 | 0.25 | - | 1.7 |
| Glass\| Py(5)\|W(3) | - | - | - | 2.2 |
| Glass\| Py(5)\|Ti(3) | - | - | - | 1.5 |
| Glass\| Py(5)\|Pt(3) | - | - | - | 2.5 |
| Glass\| Py(5)\|W(20) | - | - | - | 5.3 |
| Glass\| Py(5) | - | - | - | 2.4 |
| Glass\| Py(5)\|Ti(20) | - | - | - | 1.2 |
| Glass\| Ni(5)\|Ti(3)\|Cu(2) | 0.53 | - | - | 3.1 |
| Glass\| Ni(5)\|Pt(3)\|Cu(2) | 0.54 | - | - | 4.2 |
| Glass\| Ni(5)\|W(3)\|Cu(2) | 0.58 | - | - | 4.0 |
| Glass\| Ni(5)\|Cu(2) | 0.52 | - | - | 4.5 |
| Glass\| Ni(5)\|Cu(2)\|Ti(3) | 0.56 | - | - | 3.4 |
| Glass\| Ni(5)\|Cu(2)\|Pt(3) | 0.54 | - | - | 3.7 |
| Glass\| Ni(5)\|Cu(2)\|W(3) | 0.57 | - | - | 3.4 |
| Glass\| Ni(5)\|W(15) | 0.54 | 0.07 | 0.19 | 4.7 |
| Glass\| Ni(5)\|W(10) | 0.57 | 0.10 | 0.18 | 4.2 |
| Glass\| Ni(5)\|W(5) | 0.63 | 0.16 | 0.15 | 3.6 |
| Glass\| Ni(5)\|W(2) | 0.60 | 0.22 | 0.08 | 2.9 |
| Si\| Ni(5)\|W(3) | - | - | - | 2.9 |
| Si\| Ni(5)\|Ti(20) | - | - | - | 1.6 |
| Si\| Ti(50) | - | - | - | 1.5 |
| Si\| Ni(5)\|Pt(3) | - | - | - | 3.4 |
| Si\| Ni(5)\|W(20) | - | - | - | 4.3 |
| Si\| Ni(5) | - | - | - | 3.3 |
| Si\| Ni(5)\|Ti(3) | - | - | - | 2.3 |

**Table S1. Optical and properties of all studied samples.** To obtain the absorbed fluence in the FM and PM layer, we assume imaginary parts of the dielectric constants at 800 nm of 22.07 for Ni, 9.31 for Pt, 19.41 for Ti and 19.71 for W [58]. All films are additionally capped with 4 nm SiO$_2$. In the first column, film thicknesses in nanometers are given as numerals in parentheses.